\documentclass{aa}  

\usepackage{graphicx}
\usepackage{txfonts}
\usepackage{bm}
\usepackage[breaklinks=true]{hyperref}
\usepackage{xcolor}
\hypersetup{
    colorlinks,
    linkcolor={blue!50!black},
    citecolor={blue!50!black},
    urlcolor={blue!50!black}
}
\usepackage{natbib}
\bibpunct{(}{)}{;}{a}{}{,}

\newcommand{\msun}{~{\rm M}_{\sun}}
\newcommand{\am}{angular momentum}
\newcommand{\somedot}[2]{\dot{#1}_{\rm #2}}
\newcommand{\minsub}[2]{#1_{\rm #2}}

\begin{document}

   \title{Supermassive black hole spin evolution in cosmological simulations with \textsc{OpenGadget3}}


   \author{Luca Sala\inst{1}\thanks{email: lsala@usm.lmu.de}
          \and
          Milena Valentini\inst{2}\fnmsep\inst{3}\fnmsep\inst{1}\fnmsep\inst{4}
          \and
          Veronica Biffi\inst{3}
          \and
          Klaus Dolag\inst{1}\fnmsep\inst{5}
          }
   \institute{Universit\"ats-Sternwarte, Fakult\"at f\"ur Physik, Ludwig-Maximilians-Universit\"at M\"unchen, Scheinerstr. 1, D-81679 München, Germany
         \and
             Astronomy Unit, Department of Physics, University of Trieste, via Tiepolo 11, I-34131 Trieste, Italy
         \and
             INAF – Osservatorio Astronomico di Trieste, via Tiepolo 11, I-34131 Trieste, Italy
         \and
             ICSC - Italian Research Center on High Performance Computing, Big Data and Quantum Computing
         \and
             Max-Planck-Institut f\"ur Astrophysik, Karl-Schwarzschild-Str. 1, D-85741 Garching, Germany
             }

   \date{Received ; accepted .}

\abstract
{Mass and spin of massive black holes (BHs) at the centre of galaxies evolve due to gas accretion and mergers with other BHs. Besides affecting e.g. the evolution of relativistic jets, the BH spin determines the efficiency with which the BH radiates energy.}
{Using cosmological, hydrodynamical simulations, we investigate the evolution of the BH spin across cosmic time and its role in controlling the joint growth of supermassive BHs and their host galaxies. }
{We implement a sub-resolution prescription that models the BH spin, accounting for both BH coalescence and misaligned accretion through a geometrically thin, optically thick disc. We investigate how BH spin evolves in two idealised setups, in zoomed-in simulations, and in a cosmological volume. The latter simulation allows us to retrieve statistically robust results as for the evolution and distribution of BH spins as a function of BH properties.}
{We find that BHs with $\minsub{M}{BH}\lesssim 2 \times 10^{7}\msun$ grow through gas accretion, occurring mostly in a coherent fashion that favours spin-up. Above $\minsub{M}{BH}\gtrsim 2 \times 10^{7}\msun$ the gas \am{} directions of subsequent accretion episodes are often uncorrelated with each other. The probability of counter-rotating accretion and hence spin-down increases with BH mass. In the latter mass regime, BH coalescence plays an important role. The spin magnitude displays a wide variety of histories, depending on the dynamical state of the gas feeding the BH and the relative contribution of mergers and gas accretion. As a result of their combined effect, we observe a broad range of values of the spin magnitude at the high-mass end. Reorientation of the BH spin direction occurs on short timescales ($\lesssim 10$ Myr) only during highly-accreting phases ($\minsub{f}{Edd}\gtrsim 0.1$). Our predictions for the distributions of BH spin and spin-dependent radiative efficiency as a function of BH mass are in very good agreement with observations.}
{}


   \maketitle
%

\section{Introduction}

Supermassive black holes (BHs) ($10^6\lesssim \minsub{M}{BH}/\msun \lesssim 10^{10}$) at the centre of massive galaxies are observed as active galactic nuclei (AGN), if they accrete material from and release energy in their proximity.
This process, dubbed AGN feedback, is able to significantly affect the surroundings of the BHs and it is thought to play a significant role in the evolution of massive galaxies \citep[e.g.][]{Benson+2003, DiMatteo+2005}. Observationally, feedback has been broadly categorized in two modes, depending on the main channel of energy release \citep[][]{Churazov+2005, Sijacki+2007, Fabian2012}. In the so-called quasar mode, operating close to the Eddington limit, most of the energy is released in the form of radiation and winds \citep[][]{Silk&Rees1998, Fabian1999, Harrison2018}. This mode is prevalent at earlier epochs \citep[e.g.][]{Croton+2006}. A second mode occurs at highly sub-Eddington accretion rates, when energy is channelled mostly into powerful relativistic jets \citep[][]{McNamara&Nulsen2007,Cattaneo+2009, McNamara&Nulsen2012}. This mode is often referred to as maintenance or radio mode.

Rotating BHs have been advocated to have a fundamental role in the formation of jets \citep{Davis2020} and related feedback \citep{McNamara&Nulsen2012}. The spin $\minsub{J}{BH}$ of a BH with mass $\minsub{M}{BH}$ can range from 0 to the maximal value $\minsub{J}{max}=G\minsub{M}{BH}^2/c$. Thus, a BH is characterised by the dimensionless spin parameter $a=\minsub{J}{BH}/\minsub{J}{max}$. A maximally-spinning $10^9\msun$ BH can potentially provide $\sim10^{62}$ erg of rotational energy \citep{McNamara&Nulsen2012} that could be injected into the surroundings and offset cooling, once a mechanism to extract this energy is in action.
Theoretically, a viable mechanism has been suggested by \cite{Blandford&Znajek1977} \citep[see also][]{Lasota+2014}. 
This process requires poloidal magnetic fields anchored to an accretion flow, surrounding a rotating BH. The accreting matter drags the poloidal field into the BH ergosphere, where the BH rotation twists the field lines, generates a toroidal component of the field and exerts an effective pressure that accelerates the plasma \citep[][]{Tchekhovskoy2015}. Thus, the power of the jet is affected by the BH spin magnitude and the magnetic field flux threading the BH event horizon. 
The latter can be maximised in the so-called magnetically-arrested disc state. In this configuration, magnetic flux is accumulated near the event horizon, until it becomes dynamically relevant. In this state, the ratio between jet power and accretion power (i.e. $\eta=P_{\rm  jet}/(\dot{M}c^2)$) can be larger than one \citep[][]{Tchekhovskoy+2011,McKinney+2012,Narayan+2022}. This is possible by extracting rotational energy from the BH. Launching powerful jets requires the presence of an accretion flow that acts as an anchor for the magnetic field and drags it inwards. If a jet is created, it is possible to tap into the rotational energy of the BH. Conversely, it is possible to power an AGN by accretion alone (without extracting spin energy) but in this case the maximum power output can only reach few tens percent of the accretion power.

Observationally, a connection between the BH spin and the jet power has been proposed to explain the observed radio-loudness (radio-to-optical flux density ratio, \citealt{Sikora+2007}) in samples of AGN \citep[e.g.][]{Sikora+2007, Martinez-Sansigre&Rawlings2011}. \cite{Ghisellini+2010} find that the power of the most powerful jets in their sample of blazars \citep[see][for a definition]{Giommi+2012} can exceed the luminosity of the accretion disc by a factor of $\sim$10. They argue that extraction of rotational energy may provide the additional power required. They also argue that if this is the case, the observed correlation between $\minsub{P}{jet}$ and accretion disc luminosity may arise from a link between the accretion and rotational energy extraction processes (as discussed above). Finally, even though instrumental capabilities prevent a definite confirmation so far, future polarimetry measurements with the Event Horizon Telescope (EHT) could provide observational support to the process of energy extraction by jets \citep{Chael+2023}.

The direction of the BH spin also plays a key role in jet formation. Spinning BHs create an outward Poynting flux in the direction of the BH spin, through the Blandford-Znajek mechanism \citep[][]{Hawley&Krolik2006,Nakamura+2008}.
Several observations of nearby clusters exhibit the presence of multiple pairs of jet-inflated cavities, that have distinct angular orientations relative to the central galaxy \citep[e.g.][]{Forman+2007,David+2009,Sanders+2009,Fabian+2011}. This suggests multiple generations of jets propagating along different directions. The observations may be interpreted as spin re-orientation, provided that processes close to the BH manage to modify the BH spin and thus the launching direction of the jet (e.g. accretion proceeding on different planes across subsequent events). A few numerical works demostrated that it is possible to recover the observed morphologies by assuming a re-orienting jet \citep[][]{Cielo+2018,Horton+2020,Lalakos+2022}.

BH spins evolve across cosmic time due to accretion and mergers.  
Observational estimates of the BH spin parameter are therefore crucial to gain insight into its evolution. Several methods are adopted in literature to infer the BH spin of AGN (e.g. X-ray reflection methods or thermal continuum fitting, see \citealt{Reynolds2021}, for a review). Most of the measurements published to date are based on X-ray reflection spectroscopy based on Fe K lines. Recent works developed a new class of high-density disc models \citep[][]{Jiang+2019,Jiang+2022}, used to estimate spin in low-mass BHs by modelling the X-ray relativistic reflection continuum \citep[][]{Mallick+2022}. The observed jet powers can also be used to determine the spin \citep[][]{Daly2009, Daly2011, Daly2019, Daly2021}. This method is particularly useful because it can be applied to large samples of AGN exhibiting jets. 
Lastly, interferometric observations with the EHT might provide constraints on the spin of M87$^{*}$, from the circularity of the shadow \citep{Broderick+2022} or from the phase-twisting of light propagating near the BH \citep{Tamburini+2020}. The former method was only able to infer a clockwise rotation with the current observational capabilities (i.e. spin vector pointing away from Earth). The latter provided an estimate of $a=0.90\pm 0.05$ and an angle with respect to the line of sight $i=163^{\circ} \pm 2^{\circ}$.

From the theoretical standpoint, several analytical studies have been focused on investigating the mechanism driving spin evolution, that is likely linked to the coupling between the accretion disc and the BH spin \citep{Papaloizou&Pringle1983,Pringle1981,Scheuer&Feiler1996,Natarajan&Pringle1998,Natarajan&Armitage1999,King+2005,Martin+2007,Perego+2009}. 
If a geometrically thin, optically thick disc is misaligned with respect to the BH spin rotation axis, the innermost part aligns with the BH spin and is connected to the unperturbed outermost part through a smooth warp (Bardeen-Petterson effect). The flow through this region exerts a torque on the BH spin, modifying its direction. 
Recently, full general-relativistic magneto-hydrodynamic simulations performed by \cite{Liska+2019} were able to reproduce this effect in magnetised thin discs. 
The BH spin magnitude changes when matter is accreted at the innermost stable circular orbit (ISCO).

Further works have focused on building upon the analytical theory to understand the evolution of BH spin as a response of accretion and mergers, as well as the relative contribution between the two channels. 
A number of works adopted a semi-analytical treatment in a hierarchical cosmological context, with various recipes for the drivers of gas accretion and its effect on spin evolution \citep{Volonteri+2005a,Volonteri&Rees2005,Berti&Volonteri2008,Lagos+2009,Fanidakis+2011,Izquierdo-Villalba+2020,Griffin+2020}. 
Other studies focused on simulating evolutionary histories of BHs under the assumption of a fixed fuelling \am{} distribution and analysed how it affects the distribution of BH spins \citep{Volonteri+2007,King+2008,Dotti+2013,Zhang&Lu2019}. 
Going a step further, some works developed models suitable to be used on-the-fly in hydrodynamical simulations \citep[][]{Maio+2013,Fiacconi+2018}, to study isolated systems at high resolution. The model by \cite{Fiacconi+2018} was also coupled to novel feedback recipes for winds \citep[][]{Cenci+2020,Sala+2020} and jets \citep[][]{Talbot+2021}. 
Recently, using the same model, \cite{Bollati+2023} studied the dynamics of massive BH binaries in the presence of spin-dependent radiative feedback, whereas \cite{Talbot+2023} focused on the effect of jets in gas-rich galaxy mergers. While all the above works assumed a thin accretion disc, \cite{Husko+2022} developed a spin evolution model that assumes a thick disc and applied it to study jet feedback in galaxy clusters. 
\cite{Dubois+2014a} made a further step forward and developed a model suited to cosmological, hydrodynamical simulations, built upon \cite{Volonteri+2007} and \cite{King+2005,King+2008}. The model was used to perform zoom-in simulations \citep{Dubois+2014a} and later updated to include jets, with BH spin-dependent direction and power \citep{Beckmann+2019,Dubois+2021,Dong-Paez+2023}. 

Comparatively fewer works have been dedicated to the study of spin evolution in a cosmological box and using a large statistical sample of BHs. \cite{Dubois+2014} run their spin evolution model, although only in post-processing, on the output of the \textsc{Horizon-AGN} simulation. \cite{Bustamante&Springel2019} performed a set of simulations of a cosmological volume, evolving the spin on-the-fly with a similar model, although with a few differences at the sub-resolution level with respect to our implementation (see Sec.~\ref{sec:spinevol_algorithm} and \ref{sec:previous_work_comparison}).
In this work, we focus on studying supermassive BH spin evolution with a sub-resolution model that follows the latter two works. Our aim is to take full advantage from the statistically significant simulated population of BHs in a cosmological box. Using information provided by cosmological, hydrodynamical simulations, we aim to analyse in detail the coupled evolution of the spin direction and magnitude, the effect of the dynamical state of the gas fuelling the BHs and the impact of mergers. 

This paper is organised as follows. Sec.~\ref{sec:model_description} describes the theoretical basis of our model and its implementation. Sec.~\ref{sec:sim_suite} presents the suite of simulations we used to test our model and statistically study BH spins. Sec.~\ref{sec:results} presents our simulation results, whereas in Sec.~\ref{sec:discussion} we discuss their implications. In Sec.~\ref{sec:previous_work_comparison} we compare our work with previous studies in literature. In Sec.~\ref{sec:conclusion} we summarise and conclude. 

\section{The model}\label{sec:model_description}
In this Section, we present our sub-resolution model for spin evolution due to gas accretion onto BHs and mergers and its coupling to the resolved scales. We implement our model in the \textsc{TreePM} code \textsc{OpenGadget3} \citep[see][]{Groth+2023}, descendant of a non-public evolution of the \textsc{Gadget-3} code \citep[originally from][]{Springel2005}. The code features a modern smoothed particle hydrodynamics (SPH) description \citep{Beck+2016} and adopts a bias-corrected, sixth-order Wendland kernel \citep[][]{Dehnen&Aly2012} with 295 neighbours.

\subsection{BH treatment}
\label{sec:BH_in_sims}
Our sub-resolution model is built within the broader BH physics module for cosmological simulations originally described in \cite{Springel+2005a}, with modifications as in \cite{Hirschmann+2014} and \cite{Steinborn+2015}. 

Each BH in our simulations is treated as a collisionless sink particle, coupled to other particles only by gravity. We associate an SPH kernel to every BH (with smoothing length $\minsub{h}{BH}$) in a similar fashion as for the gas particles, with the same number of neighbours. The gravitational force on the BH is softened using a Plummer-equivalent softening length $\varepsilon_{\rm BH}$. Its value is specific for each simulation and stated in the corresponding subsections and in Table~\ref{tab:cosmo_simulations_summary}.
To select haloes where to seed new BH particles, we apply a FoF (\textit{Friends-of-Friends}) algorithm to the stellar particles alone. In this way, we identify stellar bulges. We adopt a linking length of $l=0.05$ and require that a halo has a stellar mass corresponding to $\minsub{M}{*,seed}$ to host a new BH seed.
We also demand that the halo has a stellar over dark matter (DM) mass fraction $\minsub{f}{*}>0.05$ and a gas over stellar mass fraction $\minsub{f}{gas}>0.1$, to ensure that the identified stellar bodies are newly-forming galaxies and not tidally stripped stellar remains. The seed is created at the position of the star particle with the largest binding energy within the group, with an initial mass $\minsub{M}{BH,seed}$. The values of $\minsub{M}{BH,seed}$ and $\minsub{M}{*,seed}$ for each simulation are stated in the corresponding subsections and in Table~\ref{tab:cosmo_simulations_summary}. Alternatively, BHs with specific properties can be inserted in the initial conditions (ICs) to study a specific configuration of an idealised system (Sec.~\ref{sec:ideal_gal_setup} and \ref{sec:ideal_merger_setup}).

We do not apply any pinning of the BHs to the position of the potential minimum within $\minsub{h}{BH}$. Besides, we do not apply any drag force from the continuous gas accretion onto the BH and keep position and velocity of the central BH when BHs merge.
We allow BHs to merge if the following criteria are met: \textit{(i)} their relative velocity is smaller than 0.5$c_{s}$, where $c_{s}$ is the sound speed of the gas, averaged in a kernel-weighted fashion; \textit{(ii)} their distance is smaller than 5 times $\varepsilon_{\rm BH}$; \textit{(iii)} the binary binding energy is smaller than $0.5c_{s}^2$ \citep[see][for further details]{Hirschmann+2014}.

Besides mergers, BHs also increase their mass\footnote{In our code BHs are characterised by two masses, a dynamical mass used to compute gravitational interactions and a physical mass used in all the BH sub-resolution models.} by accreting gas. The mass inflow rate onto the BH is computed using the Bondi--Hoyle--Lyttleton prescription \citep[hereafter Bondi;][]{Hoyle&Lyttleton1939, Bondi&Hoyle1944, Bondi1952}:
\begin{equation}
    \dot{M}_{\rm B}=\frac{4\pi \alpha_{\rm acc} G^2 M^2_{\rm BH} \langle\rho\rangle}{(\langle c_{\rm s}\rangle^2+\langle v\rangle^2)^{3/2}}, \label{eq:bondi}
\end{equation}
\noindent where $\rho$ is the gas density, $v$ is the velocity of the BH with respect to the gas and $G$ is the gravitational constant.
We compute the properties marked with $\langle\cdot\rangle$ as a kernel-weighted average over the BH neighbouring particles. We also enforce a maximum radius on $\minsub{h}{BH}$ for the computation, $\minsub{r}{acc}=100$ kpc. $\minsub{\alpha}{acc}$ is a dimensionless parameter that is often adopted to account for the detailed structure of the interstellar medium (ISM) in the vicinity of the BH (e.g. \citet{Springel+2005b, Booth&Schaye2009}; but see e.g. \citet{Valentini+2020}), typically unresolved in cosmological simulations because of the limited spatial resolution ($\sim$kpc). In our code we compute two different Bondi rates using $\minsub{\alpha}{acc,hot}=10$ and $\minsub{\alpha}{acc,cold}=100$ (unless stated otherwise) for the hot and cold gas phases\footnote{The hot phase is composed by particles with a temperature $T>2\times10^{5}$~K. The cold phase is composed by star-forming particles (see Sec.~\ref{sec:sim_suite}) and gas particles that have a temperature $T<2\times10^{5}$~K.} respectively. We assume that the accretion rate onto the BH is given by
\begin{equation}
    \dot{M}_{\mathrm{BH}}=\min \left(\dot{M}_{\mathrm{B}, \mathrm{hot}}+\dot{M}_{\mathrm{B}, \mathrm{cold}}, \dot{M}_{\mathrm{Edd}}\right) \label{eq:mdot}
\end{equation}
where $\dot{M}_{\rm Edd} = 4\pi G M_{\rm BH} m_{\rm p}/(\sigma_{\rm T}\minsub{\epsilon}{r} c)$ is the \citet{Eddington1916} accretion rate, $m_{\rm p}$ the proton mass, $\sigma_{\rm T}$ the Thomson cross section, $c$ is the speed of light and $\minsub{\epsilon}{r}$ is the radiative efficiency. The latter depends on the BH dimensionless spin parameter in our simulations, as explained in Sec.~\ref{sec:spinevol_theory}.

\subsection{Spin evolution algorithm}
\label{sec:spinevol_algorithm}

Our model is designed for simulations whose spatial resolution ranges from few tens of pc to a few kpc (see Sec.~\ref{sec:sim_suite}), several orders of magnitudes larger than the physical size of an accretion disc around a BH ($< 1$ pc, \citealt{Frank+2002}). In the approach often adopted in cosmological simulations, an accretion disc is not included in the modelling \citep[e.g.][]{Springel+2005a,Booth&Schaye2009}. In our implementation, we introduce a sub-grid accretion disc as an intermediate step in the mass transfer between the resolved scales and the BH. We then assume that the mass transfer rate from the resolved scales onto the accretion disc is equal to the mass rate from the accretion disc onto the BH. The mass rate is defined by Eqs.~\eqref{eq:bondi} and \eqref{eq:mdot}. We also assume that the gas maintains the angular momentum direction it has at the resolved scales. The inclusion of an accretion disc is necessary to model the physical effects that modify the spin due to gas accretion, as explained in the following.

\subsubsection{Accretion disc and spin evolution}
\label{sec:spinevol_theory}
We define the BH \am{} vector $\minsub{\bm{J}}{BH}$ as 
\begin{equation}
    \minsub{\bm{J}}{BH}=\minsub{J}{BH}\cdot \minsub{\bm{j}}{BH}=a\minsub{J}{max} \minsub{\bm{j}}{BH}, \label{eq:spindef}
\end{equation}
where $\minsub{\bm{j}}{BH}$ is the unit vector encoding its direction and $\minsub{J}{BH}$ is its magnitude. We define $0 \leq a \leq 1$. In what follows, a negative sign encodes counter-rotating conditions on a BH with spin parameter $a$. $\minsub{\bm{J}}{BH}$ evolves because of its interaction with the distribution of matter in a surrounding accretion disc, whose \am{} is misaligned with respect to $\minsub{\bm{J}}{BH}$. Our model assumes an optically thick, geometrically thin \cite{Shakura&Sunyaev1973} accretion disc.

If we start from a completely flat disc surrounding a spinning BH, Lense-Thirring precession forces the fluid elements close to the BH to precess and induces them to rotate in the BH equatorial plane. If viscosity is sufficiently high, the shear stresses propagate the perturbation diffusively until a warped steady state of the disc is reached \citep[][]{Bardeen&Petterson1975,Martin+2007}. The largest deviation from a flat profile -- i.e. disc annuli where the gas \am{} is misaligned with respect to both the unperturbed outer region of the disc and the BH spin -- occurs at the warp radius. The latter is defined as the radius $\minsub{R}{w}$ at which the Lense-Thirring precession timescale $\minsub{t}{LT} = R_{\rm w}^3 c^2 / (2G\minsub{J}{BH})$ \citep{Wilkins1972} is equal to the warp propagation timescale $\minsub{t}{w} \sim R_{\rm w}^2/\nu_2$ \citep{Pringle1981, Perego+2009}. $\nu_2$ is the vertical shear viscosity governing the propagation of vertical perturbations \citep[e.g.][]{Volonteri+2007,Perego+2009,Dotti+2013,Dubois+2014}. Following \cite{Dubois+2014} we compute
\begin{align}
    \frac{R_{\text {w}}}{R_{\mathrm{BH}}} \simeq 4 \times 10^{2} a^{5 / 8} M_{\mathrm{BH}, 8}^{1 / 8}\left(\frac{\minsub{f}{Edd}}{\epsilon_{\mathrm{r}, 01}}\right)^{-1 / 4}\left(\frac{\nu_{2} / \nu_{1}}{85}\right)^{-5 / 8} \alpha_{\nu_1, 01}^{-1 / 2}. \label{eq:rwarp_numeric}
\end{align}
where $M_{\mathrm{BH}, 8}=\minsub{M}{BH}/(10^8\msun)$, $\epsilon_{\mathrm{r}, 01}=\epsilon_{\mathrm{r}}/0.1$ and we define the Eddington ratio $\minsub{f}{Edd}=\somedot{M}{BH}/\somedot{M}{Edd}$. $\nu_1$ is the radial shear viscosity, responsible for the radial inward drift of gas across the disc. We also assume that the viscosity $\alpha$-parameter \citep{Shakura&Sunyaev1973} is $\alpha_{\nu_1, 01}=\alpha_{\nu_1}/0.1\equiv1$, following \cite{King+2005}, and $\minsub{\nu}{2}/\minsub{\nu}{1}=2(1+7\minsub{\alpha}{\nu_1})/(4+\minsub{\alpha^2}{\nu_1})/\minsub{\alpha^2}{\nu_1}$, following \cite{Ogilvie1999}, leading to a fiducial value for the ratio $\minsub{\nu}{2}/\minsub{\nu}{1}=85$.

\begin{figure}
    \centering
    \includegraphics[width=0.48\textwidth, trim=0 0 0 9, clip]{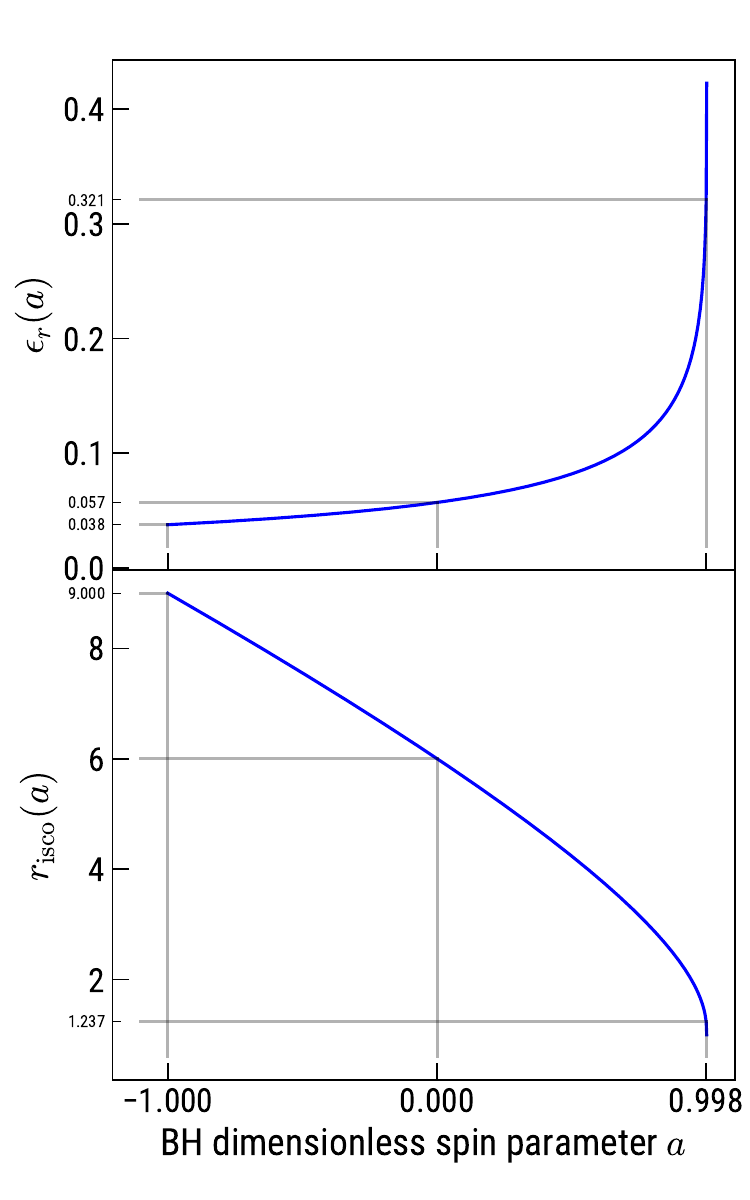}
    \caption{Radiative efficiency -- Eq.~\eqref{eq:radiative_efficiency} -- ({top panel}) and radius of the innermost stable circular orbit -- Eq.~\eqref{eq:risco} -- ({bottom panel}), as a function of the BH dimensionless spin parameter. Some reference values of these quantities are also highlighted: $a=-1$, for a counter-rotating orbit around a maximally spinning BH; $a=0$, for a non-spinning BH; $a=0.998$, for the maximum spin allowed in our simulations.}
    \label{fig:spin_funcs}
\end{figure}

The flow of matter through the warp is responsible to exert a torque that modifies \textit{only the direction} of the BH spin. Indeed, matter flowing from the unperturbed outermost region to the aligned innermost region through the warp changes its \am{} direction, therefore forcing the BH spin orientation to change and ensure conservation of the total \am{} \citep{King+2005,Dotti+2013}. 
Since the torque acting to modify the direction of the BH spin is produced by matter flowing through the warped region at $\minsub{R}{w}$, we adopt the same assumption as \cite{Volonteri+2007} and \cite{Dubois+2014} and define $\minsub{\bm{J}}{d}=\minsub{J}{d}\minsub{\bm{j}}{d}$ as the \am{} of the disc within $\minsub{R}{w}$.
\cite{King+2005} showed that starting from a configuration where $\minsub{\bm{J}}{BH}$ and $\minsub{\bm{J}}{d}$ are initially misaligned, the torque resulting from the warped configuration always leads the BH spin to align with the total \am{} $\minsub{\bm{J}}{tot}$ of the system disc+BH
\begin{align}
    \minsub{\bm{J}}{tot}&=\minsub{\bm{J}}{BH}+\minsub{\bm{J}}{d}. \label{eq:Jtot}
\end{align}
Moreover, the torque acts dissipatively on the disc, whose \am{} ends up either aligned or counter-aligned with the BH spin, depending on the ratio between the \am{} magnitudes. Namely, the counter-aligned configuration is possible if and only if 
\begin{equation}
    \cos{\theta_{\rm BH-d}} < -\frac{\minsub{J}{d}}{2\minsub{J}{BH}}\label{eq:counter-align_condition}
\end{equation} 
where $\theta_{\rm BH-d}$ is the angle subtended by the initial \am{} directions $\minsub{\bm{j}}{BH}$ and $\minsub{\bm{j}}{d}$, i.e. $\cos{\theta_{\rm BH-d}}=\minsub{\bm{j}}{BH}\cdot\minsub{\bm{j}}{d}$. 
Only the (counter-) aligned innermost -- i.e. within $\minsub{R}{w}$ -- region can effectively transfer its \am{} to the BH, once the matter enclosed within this region is eventually accreted \citep{Volonteri+2007}. Therefore, $\minsub{R}{w}$ is also the relevant radial scale to estimate the variation of the BH spin magnitude as a result of the accretion of this innermost part.
Since the radial shear viscosity $\nu_1$ is responsible for the inward drift of matter across the disc, the region within $\minsub{R}{w}$ is accreted on a timescale $\minsub{t}{\nu_1}(\minsub{R}{w})\sim R_{\rm w}^2/\nu_1$ \citep{King+2005,Perego+2009,Dubois+2014}. We follow \cite{Dubois+2014} and compute
\begin{align}
    \minsub{t}{\nu_1}(\minsub{R}{w})\sim 3.4 \times 10^{5} a^{7 / 8} M_{\mathrm{BH}, 8}^{11 / 8}\left(\frac{\minsub{f}{Edd}}{\epsilon_{\mathrm{r}, 01}}\right)^{-3 / 4}\left(\frac{\nu_{2} / \nu_{1}}{85}\right)^{-7 / 8} \alpha_{\mathrm{\nu_1{}}, 01}^{-3 / 2} \mathrm{yr}.\label{eq:tnu1}
\end{align}
Note that $\minsub{\nu}{1}\ll\minsub{\nu}{2}$ and the warp propagation timescale $\minsub{t}{w}$ is thus much shorter than the radial drift timescale $\minsub{t}{\nu_1}$. As a consequence, the timescale on which the warp forms is much shorter than that over which the spin magnitude and direction change. 
We then estimate the mass enclosed within the warped region $\minsub{M}{d}$ as
\begin{equation}
    \minsub{M}{d}\simeq\somedot{M}{BH}\minsub{t}{\nu_1}(\minsub{R}{w}).\label{eq:diskmass}
\end{equation}
where $\somedot{M}{BH}$ is given by Eq.~\eqref{eq:mdot}.
When $\minsub{M}{d}$ is accreted onto the BH, the BH spin magnitude $a$ changes due to the accretion of its ISCO \am{}, according to the expression derived in \cite{Bardeen&Petterson1975}
\begin{equation}
    \minsub{a}{f}= \frac{1}{3}\frac{\minsub{r^{1/2}}{isco}}{\minsub{M}{ratio}}\left[4-\left(3\frac{\minsub{r}{isco}}{\minsub{M^2}{ratio}}-2\right)^{1/2}\right], \label{eq:spinupdate}
\end{equation} 
where 
\begin{equation}
    \minsub{M}{ratio}=\frac{\minsub{M}{BH}^{i}+\minsub{M}{d}(1-\minsub{\epsilon}{r})}{\minsub{M}{BH}^{i}}.\label{eq:mratio}
\end{equation} 
and, normalising to the gravitational radius $\minsub{R}{g}=\minsub{R}{BH}/2=G\minsub{M}{BH}/c^2$, 
\begin{equation}
    \minsub{r}{isco}=\minsub{R}{isco}/\minsub{R}{g}=3+Z_2\pm[(3-Z_1)(3+Z_1+2Z_2)]^{1/2},\label{eq:risco}
\end{equation}
where the positive (negative) sign is for counter(co)-rotating orbits. $Z_1$ and $Z_2$ are functions of the BH dimensionless spin parameter only:
\begin{equation}
    Z_1=1+(1-a^2)^{1/3}[(1+a)^{1/3}+(1-a)^{1/3}],
\end{equation}
\begin{align}
    Z_2=(3a^2+Z_1^2)^{1/2}.
\end{align}
$\minsub{R}{isco}$ ranges from 1 to 9 $\minsub{R}{g}$, for co- and counter-rotating orbits on a maximally spinning BH respectively, as illustrated in the bottom panel of Fig.~\ref{fig:spin_funcs}. Physical sizes are therefore between $\sim 5\times10^{-6}M_{\rm BH,8}\text{ pc }$ and $\sim 5\times10^{-5}M_{\rm BH,8}\text{ pc }$. 
The mass accreted is also corrected for the radiated energy, where 
\begin{equation}
    \minsub{\epsilon}{r} = 1 - \sqrt{1-\frac{2}{3\minsub{r}{isco}}}. \label{eq:radiative_efficiency}
\end{equation}
The top panel of Fig.~\ref{fig:spin_funcs} shows the value of $\minsub{\epsilon}{r}$ as a function of $a$.

Finally, as mentioned before, we estimate the \am{} $\minsub{J}{d}$ of the accreted disc region that determines the spin variation in magnitude and direction (and is also used to evaluate condition \ref{eq:counter-align_condition}) as
\begin{equation}
    \minsub{J}{d}\simeq\minsub{M}{d}(\minsub{R}{w})\minsub{\omega}{k}(\minsub{R}{w})\minsub{R}{w}^2=\minsub{M}{d}(\minsub{R}{w})(G\minsub{M}{BH}\minsub{R}{w})^{1/2}, \label{eq:Jdisk}
\end{equation}
where $\minsub{\omega}{k}=\sqrt{(G\minsub{M}{BH}/R^3)}$ is the keplerian angular frequency.
While $\minsub{R}{isco}\sim 1-9 \minsub{R}{g}$, the warped region is at hundreds of gravitational radii (see Eq.~\ref{eq:rwarp_numeric}), thus a few orders of magnitude larger. Therefore, the BH spin direction changes on a shorter timescale than its magnitude \citep[][]{Scheuer&Feiler1996, Perego+2009, Dotti+2013}.

\subsubsection{BH spin update iteration}\label{sec:spin_udpate_iteration}
\begin{figure*}
    \centering
    \includegraphics[width=\textwidth, trim = 30 0 40 30, clip]{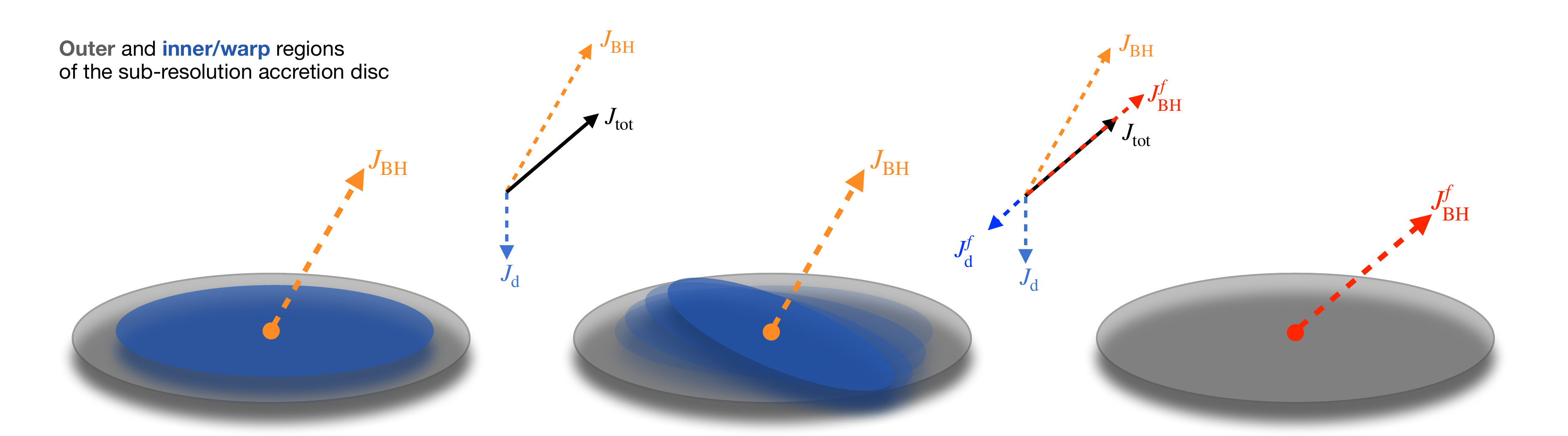}
    \caption{Schematic of the steps that compose a single accretion episode. The vector schemes in the upper part of the figure represent the initial and final configurations of the angular momenta, in a case similar to the one shown in Fig. 1b by \citet{King+2005}. \textit{Left:} an accretion disc settles around the BH, in a misaligned configuration. \textit{Center:} a warp develops, and the innermost part is forced to rotate in the BH equatorial plane and either co- or counter-align. \textit{Right:} the BH spin changes in magnitude when gas is accreted at the innermost stable orbit.}
    \label{fig:ADscheme}
\end{figure*}
The algorithm that we adopt to update BH mass and spin in our simulations models the above-mentioned processes, and makes sure that the BH mass grows consistently with the rate given by Eq.~\eqref{eq:mdot}. Following \cite{Dubois+2014}, BH mass and spin are updated in iterations (hereafter \textit{accretion episodes}) composed of the following steps: 
\begin{enumerate}
    \item we assume that an accretion disc forms around the BH, characterised by the accretion rate given by Eq.~\ref{eq:mdot}. We further assume that the initial \am{} direction of the disc $\minsub{\bm{j}}{d}$ is set by the resolved scales of the simulation, i.e. parallel to the direction of the \am{} of the gas within the BH kernel $\minsub{\bm{j}}{g}$:
    \begin{equation}
        \minsub{\bm{j}}{d}=\minsub{\bm{j}}{g}\equiv\minsub{\bm{L}}{BH,kernel}/|\minsub{\bm{L}}{BH,kernel}|,\label{eq:disc_dir}
    \end{equation}
    where
    \begin{equation}
        \minsub{\bm{L}}{BH,kernel}=\sum_{j}\minsub{m}{j}(\minsub{\bm{r}}{j}-\minsub{\bm{r}}{BH})\times(\minsub{\bm{v}}{j}-\minsub{\bm{v}}{BH})w(\minsub{\bm{r}}{j}-\minsub{\bm{r}}{BH},\minsub{h}{BH}).
    \end{equation}
    $\minsub{\bm{L}}{BH,kernel}$ is computed considering only the cold gas particles -- i.e. the component that is able to settle into an accretion disc. $w$ is the dimensionless SPH kernel function. This is the initial misaligned configuration illustrated on the left in Fig.~\ref{fig:ADscheme}.
    \item we compute $\minsub{R}{w}$ using Eq.~\ref{eq:rwarp_numeric}. This defines the region of the disc (marked in blue in Fig.~\ref{fig:ADscheme}) that exerts the alignment torque on the BH spin and whose gas is eventually accreted by the end of the accretion episode. 
    \item we compute $\minsub{M}{d}$ and $\minsub{J}{d}$, using Eq.~\ref{eq:diskmass} and \ref{eq:Jdisk} respectively. This defines the total \am{} of the accretion episode $\minsub{\bm{J}}{tot}=\minsub{\bm{J}}{BH}+\minsub{\bm{J}}{d}=\minsub{\bm{J}}{BH}+\minsub{{J}}{d}\minsub{\bm{j}}{d}$ (black solid vector in Fig.~\ref{fig:ADscheme}). Note that we are assuming that the warped distribution that defines the innermost aligned region of the disc (central panel in Fig.~\ref{fig:ADscheme}) develops on a timescale shorter than those over which the BH spin direction and magnitude change, as explained in Sec.~\ref{sec:spinevol_theory}. \label{item:disc_props}
    \item we establish whether the innermost part of the disc is co- or counter-rotating using Eq.~\eqref{eq:counter-align_condition}, by computing
    \begin{align}
            \frac{J_{\mathrm{d}}}{2J_{\mathrm{BH}}} &\simeq \frac{M_{\mathrm{d}}\left(R_{\text {w }}\right)}{a M_{\mathrm{BH}}}\left(\frac{R_{\text {w }}}{R_{\mathrm{g}}}\right)^{1 / 2}\label{eq:JdJBHratio}\\
            &\sim 6.8 \times 10^{-2} a^{3 / 16} M_{\mathrm{BH}, 8}^{23 / 16} \left(\frac{\minsub{f}{Edd}}{\epsilon_{\mathrm{r}, 01}}\right)^{1 / 8} \left(\frac{\nu_{2} / \nu_{1}}{85}\right)^{-19 / 16} \alpha_{\minsub{\nu}{1}, 01}^{-7 / 4}. \label{eq:JdJBHratio_numeric}
    \end{align}
    In case of counter-rotating conditions, $\minsub{r}{isco}$ and $\minsub{\epsilon}{r}$ are computed with a negative sign in front of $a$. \label{item:risco}
    \item the final BH spin direction changes as a result of the alignment torque and ends up parallel to the total \am{} (vector scheme on the right in Fig.~\ref{fig:ADscheme}), i.e. $\minsub{\bm{J}}{BH}^{f}\parallel\minsub{\bm{J}}{tot}$. \label{item:final_spin_dir}
    \item the disc within $\minsub{R}{w}$ is consumed and the BH spin magnitude changes according to Eq.~\eqref{eq:spinupdate} (right panel in Fig.~\ref{fig:ADscheme}), with $\minsub{M}{d}$ computed in step \ref{item:disc_props}. Counter-alignment is taken into consideration as described in step \ref{item:risco}. We also cap the spin parameter to 0.998, which is the maximum spin allowed if photon trapping is assumed \citep{Thorne1974}. The BH mass is increased by $\Delta M_{\rm BH}=\minsub{M}{d}(1-\minsub{\epsilon}{r})$.
\end{enumerate}

We stress that it is possible to update the BH magnitude and direction separately because the latter changes over a shorter timescale than the former, i.e. first the BH spin aligns with the total \am{}, then the magnitude changes because of accretion at the ISCO. Moreover, Eq.~\eqref{eq:spinupdate} does not depend on the direction, because it models \am{} accretion from the innermost disc region, that is (counter-)aligned with the equatorial plane of the spinning BH due to the Bardeen-Petterson effect.

We also note that the mass per accretion episode $\minsub{M}{d}$ might be smaller or larger than the amount of mass required to be accreted during the simulation timestep $\Delta t$. Therefore, we follow \citet{Bustamante&Springel2019} and adopt the following strategy:
\begin{itemize}
    \item if $M_{d}<\dot{M}_{\rm BH}\Delta t$, multiple accretion episodes occur over one BH simulation timestep $\Delta t$. We allow for timestep sub-cycles indexed with a counter variable, therefore executing $N=\dot{M}_{\rm BH}\Delta t/M_{d}$ accretion episodes. All the sub-cycles share the same accretion rate, computed at the beginning of the timestep. At the end of each sub-cycle (i.e. accretion episode) we update BH spin and mass.
    \item if $M_{d}>\dot{M}_{\rm BH}\Delta t$, we adopt the same strategy extending the counter across multiple timesteps. The code executes $N=M_{d}/(\dot{M}_{\rm BH}\Delta t)$ timesteps, then the accretion episode ends and the BH mass and spin are updated using averages for the accretion rate necessary in Eqs.~\eqref{eq:rwarp_numeric}, \eqref{eq:diskmass},  and \eqref{eq:JdJBHratio}, i.e.
    \begin{equation}
        \langle\somedot{M}{BH}\rangle_{t}=\sum_{i=1}^{N}\dot{M}_{\mathrm{BH},i}\Delta t_{i}\Big/\sum_{i=1}^{N}\Delta t_{i}, 
    \end{equation}
\end{itemize}

\begin{figure}
    \centering
    \includegraphics[width=0.48\textwidth, trim=0 0 0 0, clip]{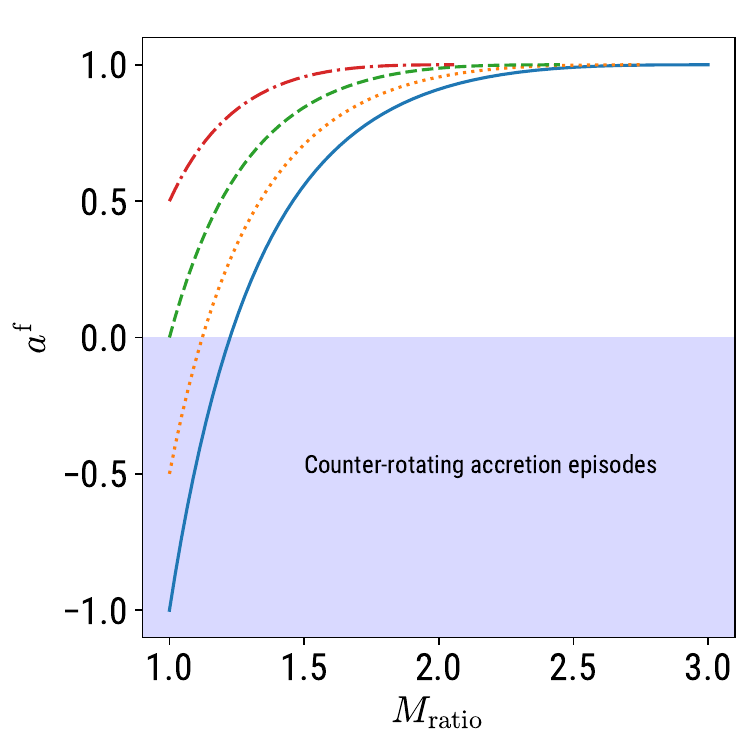}
    \caption{Final BH spin dimensionless parameter after a single accretion episode as a function of $\minsub{M}{ratio}$ as defined in Eq.~\eqref{eq:mratio}. The solid, dotted, dashed and dot-dashed lines represent $a^{\rm f}$ for the initial spin values -1, -0.5, 0, and 0.5 respectively. The solid and dotted lines represent the event of an accretion episode in which the accretion disc is counter-rotating with respect to the BH.
    }
    \label{fig:af_mratio}
\end{figure}

Fig.~\ref{fig:af_mratio} illustrates how the final value of $a$ after a \textit{single accretion episode} depends on $\minsub{M}{ratio}$, for a few example values of initial $a$, assuming no misalignment is present. Note that the plotted lines do not correspond to evolutionary tracks of the BH in our simulations, the latter ones being defined instead by a succession of multiple accretion episodes, each one characterised by different initial directions and different values for $\minsub{M}{ratio}$ and $\minsub{r}{isco}$. The solid and dotted lines correspond to counter-rotating events. Fig.~\ref{fig:af_mratio} shows that a single counter-rotating accretion episode on a maximally spinning BH (solid line) would be able to spin it down to $a=0$, if the accreted mass per episode were $\simeq25$\% of the BH initial mass. It would require an accreted mass of $\simeq2$ times the BH mass to spin it up to $a=1$, in a direction opposite to the initial one. Similarly, a counter-rotating accretion episode on a BH with $a=0.5$ (dotted line) would require $\simeq0.1\minsub{M}{BH}$ and $1.75\minsub{M}{BH}$ to achieve the same results, respectively. An accretion episode on a non-spinning BH (dashed line) needs to be $1.5\minsub{M}{BH}$ to spin it up to its maximal value, whereas it needs to be equal to $\simeq\minsub{M}{BH}$ to obtain the same result on a BH with $a=0.5$ in co-rotating conditions (dash-dotted line).

\subsubsection{Self-gravity regime}

Depending on the physical conditions, parts of the accretion disc could become unstable because of their own self-gravity \citep{Dotti+2013}. This occurs at radii that are beyond
\begin{equation}
    \frac{R_{\mathrm{sg}}}{R_{\mathrm{BH}}} \simeq 5 \times 10^{2}  M_{\mathrm{BH}, 8}^{-52 / 45}\left(\frac{\minsub{f}{Edd}}{\epsilon_{\mathrm{r}, 01}}\right)^{-22 / 45}\alpha_{\minsub{\nu}{1}, 01}^{28 / 45}.\label{eq:selfgravradius}
\end{equation}
The mass stable against fragmentation \cite[see][]{Dotti+2013} is then 
\begin{equation}
    M_{\mathrm{sg}} \simeq 6 \times 10^{5} M_{\mathrm{BH}, 8}^{34 / 45}\left(\frac{\minsub{f}{Edd}}{\epsilon_{\mathrm{r}, 01}}\right)^{4 / 45} \alpha_{\mathrm{\nu_1}, 01}^{-1 / 45}\mathrm{M}_{\odot}.\label{eq:selfgravmass}
\end{equation}
Under this condition, only the region of the disc within $\minsub{R}{sg}$ can be accreted. Therefore, if $\minsub{R}{sg}<\minsub{R}{w}$ we set $\minsub{M}{d}=\minsub{M}{sg}$ and substitute $\minsub{R}{w}$ with $\minsub{R}{sg}$ in Eqs.~\eqref{eq:Jdisk} and \eqref{eq:JdJBHratio}. In this case, 
\begin{align}
        \frac{J_{\mathrm{d}}}{2J_{\mathrm{BH}}}
        &\simeq 9.4 \times 10^{-2} a^{-1} M_{\mathrm{BH}, 8}^{-37 / 45} \left(\frac{\minsub{f}{Edd}}{\epsilon_{\mathrm{r}, 01}}\right)^{- 7 / 45} \alpha_{\minsub{\nu}{1}, 01}^{13 / 45}. \label{eq:JdJBHratio_sg_numeric}
\end{align}

Whenever a BH is seeded, its spin is set to zero. As soon as $\somedot{M}{BH}>0$, a disc is initialised with an angular momentum equal to 
\begin{equation}
    \minsub{J}{d}=\minsub{M}{sg}(G\minsub{M}{BH}\minsub{R}{sg})^{1/2},
\end{equation}
since we assume that only the portion of the disc that is stable against fragmentation can eventually accrete on the BH. The rest of the algorithm proceeds as before. $\minsub{M}{sg}$ and $\minsub{R}{sg}$ are computed using Eqs.~\eqref{eq:selfgravradius} and \eqref{eq:selfgravmass} respectively.

\subsection{BH mergers}
\label{sec:bh_mergers}

We implement a prescription to account for spin evolution in the case of BH mergers. In what follows, we adopt the same strategy as \citet{Dubois+2014} and \citet{Bustamante&Springel2019}. We use equations retrieved in a full general-relativistic framework by \citet{Rezzolla+2008} to compute the final spin after a merger event. Following their notation, we define $\bm{a}=a\minsub{\bm{j}}{BH}$. Once two BHs characterised by masses $\minsub{M}{1}$, $\minsub{M}{2}$ and spins $\minsub{a}{1}$, $\minsub{a}{2}$ have been selected to merge according to the criteria described in Sec.~\ref{sec:BH_in_sims}, the final spin vector of the remnant BH $\bm{a}^f$ is given by 
\begin{equation}
    \bm{a}^f=\frac{1}{(1+q)^2}\left(\bm{a}_1+\bm{a}_2 q^2+\bm{\ell} q\right) ,
    \label{eq:merger_a}
\end{equation}
where $q=M_2/M_1$, with $M_1\geq M_2$. In this formula, $\minsub{M}{1}$ and $\minsub{M}{2}$ are the BH physical masses. $\bm{\ell}=\bm{\ell}'/(M_1 M_2)$ where $\bm{\ell}'$ is the binary orbital angular momentum that cannot be radiated away in gravitational waves before coalescence. 
The magnitude of $\bm{\ell}$ is provided by \citet{Rezzolla+2008} and reads
\begin{equation}
    \begin{aligned}
        \ell= & \frac{s_4}{\left(1+q^2\right)^2}\left(a_1^2+a_2^2 q^4+2 \bm{a}_1 \cdot \bm{a}_2 q^2\right) \\
        & +\left(\frac{s_5 \mu+t_0+2}{1+q^2}\right)\left(a_1 \cos \phi_1+a_2 q^2 \cos \phi_2\right) \\
        & +2 \sqrt{3}+t_2 \mu+t_3 \mu^2.
    \end{aligned}
\end{equation}
Here, $\cos{\phi}=\bm{a}\cdot\bm{\ell}/(a\ell)$ depicts the
angle subtended by the spin vector of BH with $\bm{\ell}$, $\mu=q/(1+q)^2$. $s_4=-0.129, s_5=-0.384, t_0=-2.686, t_2=-3.454$, $t_3=2.353$ are the parameters of the fit to their numerical results.
We assume that $\bm{\ell}$ is parallel to the binary \am{} $\bm{L}=\bm{L}_1+\bm{L}_2$ \citep{Rezzolla+2008}. $\bm{L}_{i=1,2}$ is the \am{} vector of BH $i$ with respect to the binary centre of mass (CM), computed as $\bm{L}_i=M_i(\bm{r}_i-\bm{r}_{\rm CM})\times(\bm{v}_i-\bm{v}_{\rm CM})$. The values for the radii $\bm{r}_i$ and velocities $\bm{v}_i$ are retrieved when the BHs are selected to merge.

\subsection{AGN feedback}
\label{sec:agn_feedback}
Our AGN feedback model assumes that a fraction $\minsub{\epsilon}{f}$ of the radiated energy is coupled to the local ISM (refer to Table~\ref{tab:cosmo_simulations_summary} and following sections for values), as often adopted in cosmological simulations \citep[see e.g.][]{Springel+2005b,Booth&Schaye2009}.
The total rate of coupled energy is thus:
\begin{equation}
    \dot{E}=\minsub{\epsilon}{f}\minsub{\epsilon}{r}\dot{M}_{\rm BH}c^2,
    \label{eq:feed_energy}
\end{equation}
where $\minsub{\epsilon}{f}$ is the coupling efficiency of the feedback energy to the surrounding ISM.
An amount of energy $\Delta E=\dot{E}\Delta t$ is distributed among the particles surrounding each BH in a weighted fashion, using the SPH kernel. Moreover, we follow \cite{Hirschmann+2014} and implement a transition from quasar- to maintenance-mode feedback by assuming a coupling efficiency $\minsub{\epsilon}{f}$ larger by a factor of 4 when $\minsub{f}{Edd}<0.01$. In our model, the radiative efficiency $\minsub{\epsilon}{r}$ depends on $a$ through Eq.~\eqref{eq:radiative_efficiency}. This is at variance with other implementations where $\minsub{\epsilon}{r}$ is either kept fixed \citep[e.g.,][]{Springel+2005a, Booth&Schaye2009, Hirschmann+2014} or depends on the accretion rate and BH mass using a phenomenological prescription \citep[e.g.,][]{Steinborn+2015}.

\section{The suite of simulations}\label{sec:sim_suite}

In the following sections we present our suite of simulations. 
The simulations are carried out with an advanced formulation of SPH \citep{Beck+2016}, including a low-viscosity scheme \citep[][]{Dolag+2005,Donnert+2013} and isotropic thermal conduction \citep{Dolag+2004}. The code also features metallicity-dependent radiative cooling, using the procedure described in \cite{Wiersma+2009}, and accounts for the presence of a uniform, time-dependent ionising background \citep{Haardt&Madau2001}. We adopt the model by \cite{Springel&Hernquist2003} to include a stochastic treatment of star formation and to describe the multiphase ISM, so that each gas particle (with hydrogen number density above a threshold $n_{\rm H}=0.5\;\rm cm^{-3}$) is made up of a cold, star-forming phase in pressure equilibrium with a hot phase. A detailed chemical evolution model \citep{Tornatore+2007} allows us to account for the metal enrichment of the ISM by ageing and exploding stars.
We assume the lifetime function by \cite{Padovani&Matteucci1993}, and stellar yields for supernovae type Ia by \cite{Thielemann+2003}, supernovae type II by \cite{Nomoto+2013} and asymptotic giant branch stars by \cite{Karakas&Lattanzio2007}. Each star particle represents a simple stellar population described by a \cite{Chabrier2003a} initial mass function.

We consider two idealised cases, a galaxy in isolation (Sec.~\ref{sec:ideal_gal_setup}) and a galaxy merger (Sec.~\ref{sec:ideal_merger_setup}), to test the model in a simple setup and study the evolution of the BH spin in a well-controlled environment. We also consider three setups in a full cosmological context (Sec.~\ref{sec:zoomin_setup} and \ref{sec:cosmo_setup}), which is key to capture the effects that the complex interplay between accretion and feedback can have on BH spin evolution. All the tests are performed including cooling, as well as star formation and evolution. The cosmological simulations include AGN and stellar feedback, whereas the idealised tests are performed with stellar feedback but no AGN feedback.

\begingroup
\renewcommand{\arraystretch}{1.3}
\begin{table}
\caption{Parameter summary for the idealised tests. $\minsub{M}{BH,0}$ is the BH mass in the ICs, $\theta_{z,0}$ is the angle subtended by the initial BH spin direction and the positive z axis, $\minsub{f}{Edd}=\somedot{M}{BH}/\somedot{M}{Edd}$ is the Eddington ratio.}
\centering
\label{tab:isol_simulations_summary}
\begin{tabular}{llll}
\multicolumn{4}{c}{\textbf{Idealised Milky Way galaxy}}                                                                \\ \hline  \hline
\multicolumn{1}{c}{}  & $\theta_{z,0}\;\rm[^\circ]$     & $\minsub{M}{BH,0}\;\rm [M_{\odot}]$  &   $\minsub{f}{Edd}$   \\ \hline
IdealGal-fid             & 170                             & $5\times 10^6$                       &   1   \\
IdealGal-120             & 120                             & $5\times 10^6$                       &   1   \\
IdealGal-30              & 30                              & $5\times 10^6$                       &   1   \\
IdealGal-5e5             & 170                             & $5\times 10^5$                       &   1   \\
IdealGal-5e7             & 170                             & $5\times 10^7$                       &   1   \\
IdealGal-0.3             & 170                             & $5\times 10^6$                       &   0.3   \\
\hline
\end{tabular}
\end{table}
\endgroup

\subsection{Idealised Milky Way galaxy}\label{sec:ideal_gal_setup}
The first test we present is based upon the setup described in \cite{Steinwandel+2019} aimed at modelling a Milky Way-like galaxy with total halo mass of $M_{200}=10^{12}\;\rm\minsub{M}{\odot}$ in isolation with a BH at the potential minimum. $M_{200}$ is the mass enclosed within the spherical region whose average density is 200 times the critical density of the Universe. We defer the reader to their paper for the detailed procedure used to generate the ICs.

Table~\ref{tab:isol_simulations_summary} summarises the initial properties of the central BH in our tests. We consider a fiducial case (IdealGal-fid), initialised with $\minsub{M}{BH,0}=5\cdot10^{6}\msun$, $\minsub{a}{0}=0.5$, $\minsub{\theta}{z,0}=170^{\circ}$ and $\minsub{f}{Edd}=1$. $\minsub{\theta}{z}$ is the angle subtended by the BH spin and the z-axis. We perform further tests modifying one parameter at a time with respect to the fiducial run. Each simulation is labelled according to the value of the modified parameter, as in Table~\ref{tab:isol_simulations_summary}.
Each test assumes a fixed Eddington fraction $\minsub{f}{Edd}$ for the duration of the simulation. 

The resolution is the same for all the tests. Masses of DM, gas and star particles are: $\minsub{m}{DM}=9.6\times10^5\;\rm\minsub{M}{\odot}$, $\minsub{m}{g}=\minsub{m}{*}=4.8\times10^4\;\rm\minsub{M}{\odot}$, respectively; the softening lengths are: $\minsub{\varepsilon}{DM}=218\;\rm pc$ and $\minsub{\varepsilon}{g}=\minsub{\varepsilon}{*}=\minsub{\varepsilon}{BH}=50\;\rm pc$.

\subsection{Idealised galaxy merger} \label{sec:ideal_merger_setup}
The second set of ICs describes a galaxy merger with a mass ratio of 1:1, initialised following the procedure described in \cite{Karademir+2019}. 

\begin{figure}
    \centering
    \includegraphics[width=0.48\textwidth, trim=0 0 0 0, clip]{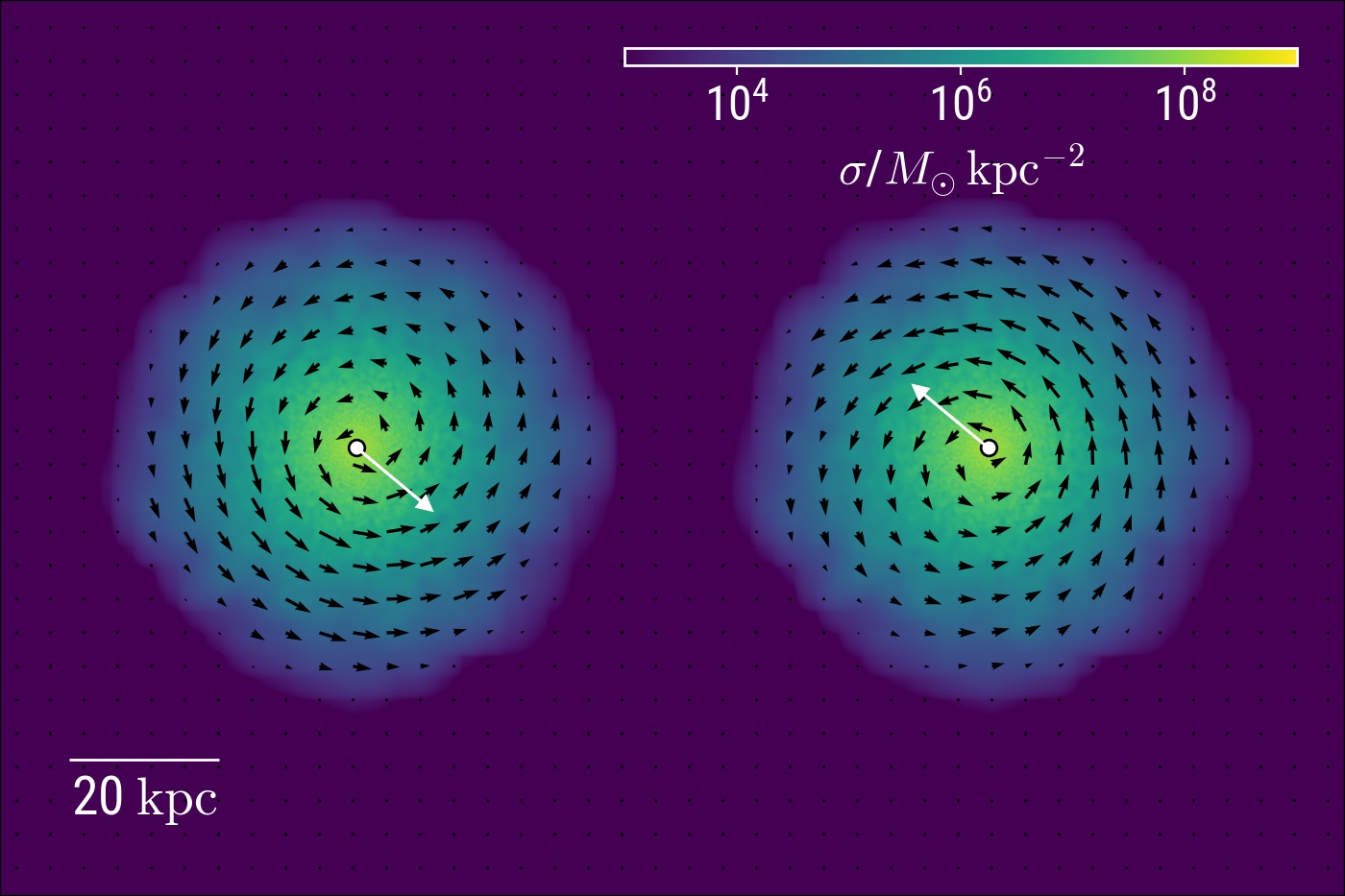}
    \caption{Gas density map of the ICs for the galaxy merger. The black arrows trace the local gas projected velocity field, while the white arrows indicate the initial direction of the CM of each galaxy. The arrows are scaled arbitrarily for visualisation purposes.}
    \label{fig:merger_ic_map}
\end{figure}

The setup consists of two identical Milky Way-like galaxies, each with mass $\minsub{M}{200}=1.89\cdot10^{12}\;\rm M_{\odot}$. Each of them is embedded in a DM halo and has a spherical stellar bulge component, as well as an exponential stellar and gas discs. The initial separation of their CMs is 80 kpc and they rotate one around the other in the same direction in which they complete their revolution. The angle between the line connecting the two galaxy centres and their initial velocity vectors is equal to $40^{\circ}$. Fig.~\ref{fig:merger_ic_map} shows the setup. The BHs are initialised with $a=0$ and $\minsub{M}{BH}=2\times10^{5}\;\rm M_{\odot}$.

In this setup, $\minsub{m}{DM}=3.22\times10^6\;\rm\minsub{M}{\odot}$, $\minsub{m}{g}=\minsub{m}{*}=1.86\times10^6\;\rm\minsub{M}{\odot}$, $\minsub{\varepsilon}{DM}=83\;\rm pc$ and $\minsub{\varepsilon}{g}=\minsub{\varepsilon}{*}=\minsub{\varepsilon}{BH}=20\;\rm pc$.

\begingroup
\renewcommand{\arraystretch}{1.5}

\begin{table*}
\caption{
Parameter summary of the cosmological simulations. $\minsub{a}{seed}$ is the BH spin parameter at seeding;
$\epsilon_f$ is the feedback coupling efficiency (Sec.~\ref{sec:agn_feedback});
$\minsub{M}{BH,seed}$ is the BH mass at seeding;
$\minsub{M}{*,seed}$ is the stellar mass considered for BH seeding (Sec.~\ref{sec:BH_in_sims});
$\minsub{m}{DM}$ is the DM particle mass;
$\minsub{m}{g}$ is the initial gas particle mass;
$\minsub{m}{*}$ is the star particle mass;
$\minsub{\varepsilon}{DM}$, $\minsub{\varepsilon}{g}$, and $\minsub{\varepsilon}{*}$ are the DM, gas, and stellar and BH softening length, respectively.
}
\centering
\begin{tabular}{lllllllllll}
 &
  $\minsub{a}{seed}$ &
  $\epsilon_f$ &
  $\minsub{M}{BH,seed}$ &
  $\minsub{M}{*,seed}$ &
  $\minsub{m}{DM}$ &
  $\minsub{m}{g}$ &
  $\minsub{m}{*}$ &
  $\minsub{\varepsilon}{DM}$ &
  $\minsub{\varepsilon}{g}$ &
  $\minsub{\varepsilon}{*,BH}$ \\
 &
   &
   &
  $\rm [M_{\odot}/h]$ &
  $\rm [M_{\odot}/h]$ &
  $\rm [M_{\odot}/h]$ &
  $\rm [M_{\odot}/h]$ &
  $\rm [M_{\odot}/h]$ &
  $\rm [kpc/h]$ &
  $\rm [kpc/h]$ &
  $\rm [kpc/h]$ \\ \hline
\multicolumn{11}{c}{\textbf{Zoom-in simulations}} \\ \hline \hline
\textsc{asin}/\textsc{dfrogin} &
  0 &
  0.05 &
  $2\times 10^{5}$ &
  $1\times10^{9}$ &
  $3.3\times 10^7$ &
  $6.25\times 10^7$ &
  $1.6\times 10^7$ &
  1 &
  1 &
  0.25 \\
\multicolumn{11}{c}{\textbf{Cosmological box}} \\ \hline \hline
\textsc{Box4} &
  0 &
  0.0775 &
  $4\times 10^{5}$ &
  $1.6\times10^{10}$ &
  $6.9\times 10^8$ &
  $1.4\times 10^8$ &
  $3.5\times 10^7$ &
  3.75 &
  3.75 &
  2.0
\end{tabular}
\label{tab:cosmo_simulations_summary}
\end{table*}
\endgroup

\subsection{Zoom-in simulations}\label{sec:zoomin_setup}

We perform our zoom-in simulations starting from ICs generated with the procedure described in \cite{Bonafede+2011}. We consider two halos: a region with target mass $M_{\rm 200}\simeq10^{12}\;h^{-1}\;\rm M_{\odot}$, dubbed \textsc{asin}; a region with target mass $M_{\rm 200}\simeq10^{13}\;h^{-1}\;\rm M_{\odot}$, dubbed \textsc{dfrogin}. They are selected from a parent DM-only simulation of a periodic box with side length $\minsub{L}{box}=1\;h^{-1}\;\rm{cGpc}$ and resolution $\minsub{m}{DM}= 10^{9}\;h^{-1}\;\msun$. 
The final resolution is quoted in Table~\ref{tab:cosmo_simulations_summary}, together with the properties and parameters of each simulation. This set of simulations adopts a flat $\Lambda$CDM cosmological model, with a matter density parameter $\minsub{\Omega}{m} = 0.24$, a baryon density parameter $\minsub{\Omega}{b}= 0.04$ and $h = 0.72$. The initial power spectrum follows $n = 0.96$ and is normalised to $\minsub{\sigma}{8} = 0.8$.

\subsection{Cosmological box}\label{sec:cosmo_setup}

We carry out a simulation starting from the ICs of the cosmological volume with $\minsub{L}{box}=48\;h^{-1}\;\rm cMpc$ (hereafter Box4) of the Magneticum\footnote{\href{http://www.magneticum.org}{http://www.magneticum.org}} simulations suite. Such a simulation, that contains $\sim 10^3$ BHs at redshift $z=0$, allows us to analyse the properties of a statistically significant BH population, rather than focusing only on the specific behaviour of a few BHs.

Table~\ref{tab:cosmo_simulations_summary} summarises the properties and parameters of the simulation. It assumes a flat $\Lambda$CDM model, with $\minsub{\Omega}{m}=0.272$, $\minsub{\Omega}{b}= 0.0456$ and $h = 0.704$, $n = 0.963$ and $\minsub{\sigma}{8} = 0.809$.

\section{Results}\label{sec:results}

\subsection{Idealised Milky Way galaxy}\label{sec:ideal_gal_results}
\begin{figure}
    \centering
    \includegraphics[width=0.48\textwidth, trim=0 0 0 0, clip]{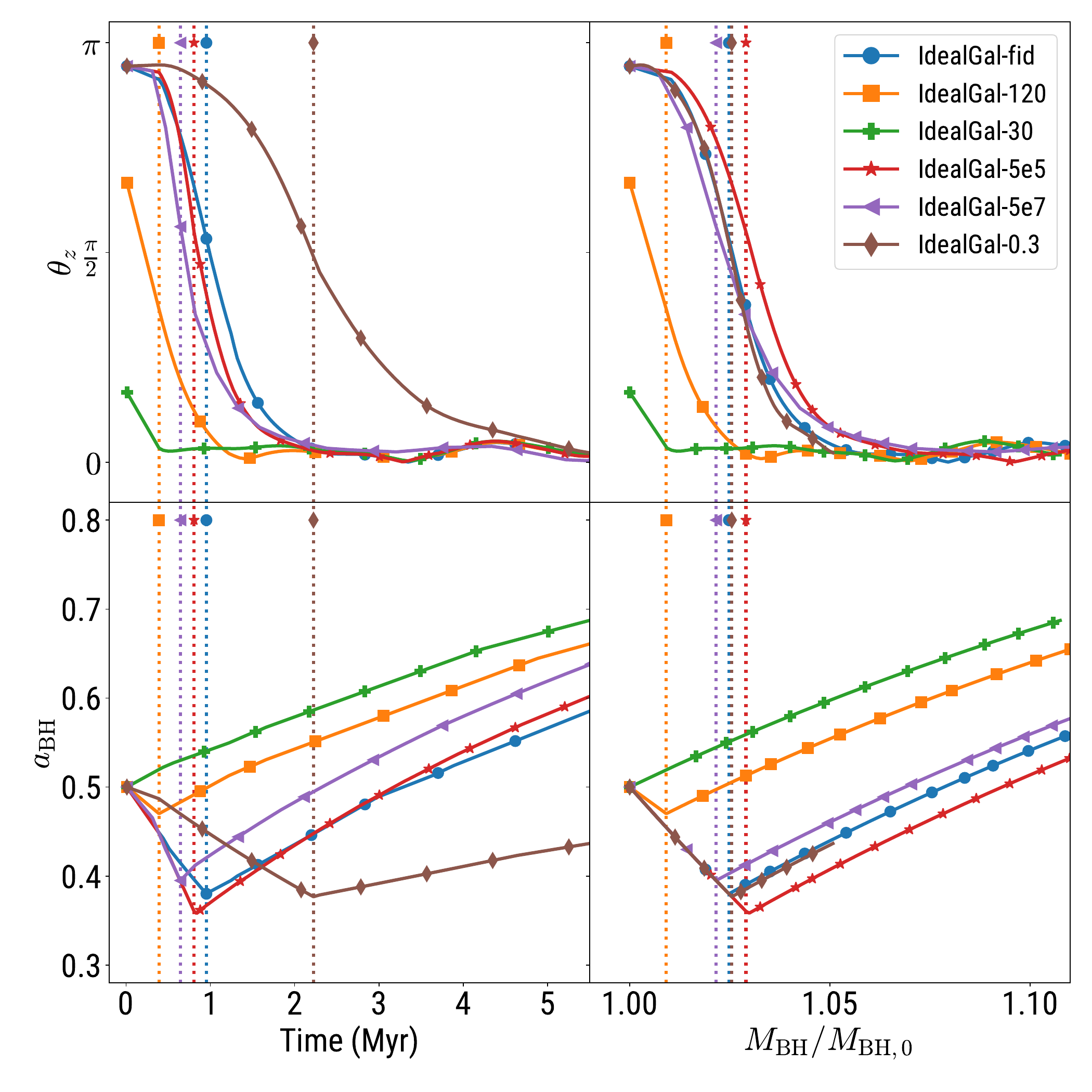}
    \caption{Spin alignment process in the idealised Milky-Way galaxy. Top panels: angle subtended by the BH angular momentum and the z-axis $\minsub{\theta}{z}$. Bottom panels: BH spin parameter $a$. Quantities are plotted as a function of time (left panels) and of the ratio between accreted and initial BH mass $\minsub{M}{BH}/\minsub{M}{BH,0}$ (right). Each line represents one simulation of the set summarised in Table~\ref{tab:isol_simulations_summary}. Dotted lines represent the instant after which condition \eqref{eq:counter-align_condition} is no longer satisfied.}
    \label{fig:isol_gal_accr}
\end{figure}
Fig.~\ref{fig:isol_gal_accr} presents the BH spin evolution in the tests listed in Table~\ref{tab:isol_simulations_summary}.
Top left panel shows the time  evolution of $\minsub{\theta}{z}$, the angle between $\minsub{\bm{j}}{BH}$ and the unit vector indicating the positive z-axis $\minsub{\bm{j}}{z}$.
The isolated galaxy is initialised with the \am{} of the accreting gas along $\minsub{\bm{j}}{z}$ (i.e. $\minsub{\theta}{z}=\pi$ refers to the BH spin direction being anti-parallel to the galaxy \am{}, while $\minsub{\theta}{z}=0$ refers to complete alignment). According to condition \eqref{eq:counter-align_condition}, counter-rotating accretion requires both $\minsub{\bm{j}}{d}\cdot\minsub{\bm{j}}{BH}>\pi/2$ and $\minsub{J}{d}<2\minsub{J}{BH}$. In IdealGal-fid, IdealGal-120, IdealGal-5e5, IdealGal-5e7 and IdealGal-0.3 both conditions are satisfied at the beginning. However, the accreting gas in the BH surroundings keeps the same direction across the entire simulation, i.e. $\minsub{\bm{j}}{g}$ remains constant. As a result, the initial sub-grid disc direction $\minsub{\bm{j}}{d}$ for every accretion episode is also constant (Eq.~\ref{eq:disc_dir}, $\minsub{\bm{j}}{g}=\minsub{\bm{j}}{d}$). Therefore, each of them induces the BH spin to tilt towards the \am{} direction of the large-scale external reservoir. Note that also $\minsub{J}{d}/\minsub{J}{BH}$ is approximately constant in these runs ($a$ and $\minsub{\epsilon}{r}$ vary slightly, but $\minsub{J}{d}/\minsub{J}{BH}$ depends weakly on them, see Eq.~\ref{eq:JdJBHratio_numeric}). Whether or not condition \eqref{eq:counter-align_condition} is verified in these tests depends only on $\minsub{\bm{j}}{d}\cdot\minsub{\bm{j}}{BH}$ (actually only on $\minsub{\bm{j}}{BH}$, since $\minsub{\bm{j}}{d}$ is constant). Each vertical dotted line marks the first instant in which accretion becomes co-rotating (from counter-rotating; i.e., condition~\eqref{eq:counter-align_condition} is no longer verified after this instant).
In IdealGal-fid the BH spin completes its alignment with the external reservoir in $\sim 2$ Myr. If we keep the same Eddington ratio but change the BH mass (IdealGal-5e5 and IdealGal-5e7, respectively), alignment takes the same time. On the other hand, alignment takes longer ($\sim 5$ Myr) to complete with the same mass but a lower 
Eddington ratio (IdealGal-0.3). At fixed BH mass and Eddington ratio (IdealGal-fid, -120, and -30), the alignment timescale depends weakly on the initial misalignment angle. A smaller $\minsub{\theta}{z,0}$ leads to a faster alignment. 

The bottom left panel of Fig.~\ref{fig:isol_gal_accr} illustrates the time evolution of the BH spin parameter $a$. Since IdealGal-fid, IdealGal-120, IdealGal-5e5, IdealGal-5e7 and IdealGal-0.3 start with counter-rotating accretion conditions, $a$ decreases. At $t\simeq1$ Myr co-rotating conditions take over and $a$ starts increasing. The turnaround point occurs in correspondence of the dotted line, after which condition \eqref{eq:counter-align_condition} is no longer verified. In IdealGal-30, $a$ monotonically increases as condition \eqref{eq:counter-align_condition} is not verified from the very beginning.

The top right panel of Fig.~\ref{fig:isol_gal_accr} shows $\minsub{\theta}{z}$ as a function of the ratio $\minsub{M}{BH}/\minsub{M}{BH,0}$. At fixed $\minsub{\theta}{z,0}$ (IdealGal-fid, -5e5, -5e7, and -0.3), the same ratio of accreted mass over BH mass ($\minsub{M}{BH} \sim 5\% \, \minsub{M}{BH,0}$) is needed for complete alignment to occur, regardless of other parameters. The actual time it takes to align depends on how fast the BH accretion proceeds. BHs whose spin starts with smaller $\minsub{\theta}{z,0}$ (IdealGal-120 and -30) require less time.
 
The bottom right panel of Fig.~\ref{fig:isol_gal_accr} presents $a$ as a function of the ratio $\minsub{M}{BH}/\minsub{M}{BH,0}$. Similarly to the bottom left panel, the turnaround point occurs in correspondence of the dotted line, when accretion switches from counter- to co-rotating. This plot further shows that it occurs at about the same value of $\minsub{M}{BH}/\minsub{M}{BH,0}$ at fixed $\minsub{\theta}{z,0}$. Finally, for smaller $\minsub{\theta}{z,0}$ a smaller ratio is required to reach the turnaround point.

The behaviour of both $\minsub{\theta}{z}$ and $a$ as a function of $\minsub{M}{BH}/\minsub{M}{BH,0}$ is similar for simulations assuming the same $\minsub{\theta}{z,0}$ because $\minsub{f}{Edd}$ is fixed. Indeed, the BH grows according to the following differential equation
\begin{equation}
    \frac{\mathrm{d}\minsub{M}{BH}}{\mathrm{d}t}=(1-\epsilon_{\rm r})\minsub{f}{Edd}\somedot{M}{Edd}=\minsub{f}{Edd}\frac{1-\epsilon_{\rm r}}{\epsilon_{\rm r}}\frac{\minsub{M}{BH}}{\minsub{\tau}{S}},
\end{equation}
where $\minsub{\tau}{S}=\sigma_{\rm T}c/(4 \pi G m_{\rm p})\sim 4.5\times10^8$ yr is the Salpeter time. As a result,
\begin{equation}\label{eq:exp_growth}
    \frac{\minsub{M}{BH}}{\minsub{M}{BH,0}}=\exp{\left(\minsub{f}{Edd}\frac{1-\epsilon_{\rm r}}{\epsilon_{\rm r}}\frac{t}{\minsub{\tau}{S}}\right)}.
\end{equation}
Therefore, the evolution of $\minsub{\theta}{z}$ and $a$ as a function of time is different depending on $\minsub{f}{Edd}$ (compare IdealGal-fid and IdealGal-0.3). On the other hand, the curves overlap if we consider $\minsub{M}{BH}/\minsub{M}{BH,0}$ as the independent variable (and they are independent of $\minsub{M}{BH,0}$ -- see IdealGal-fid, IdealGal-5e5, IdealGal-5e7).

\subsection{Idealised galaxy merger}\label{sec:ideal_merger_results}

\begin{figure*}
    \centering
    \includegraphics[width=0.48\textwidth, trim=0 0 0 0, clip]{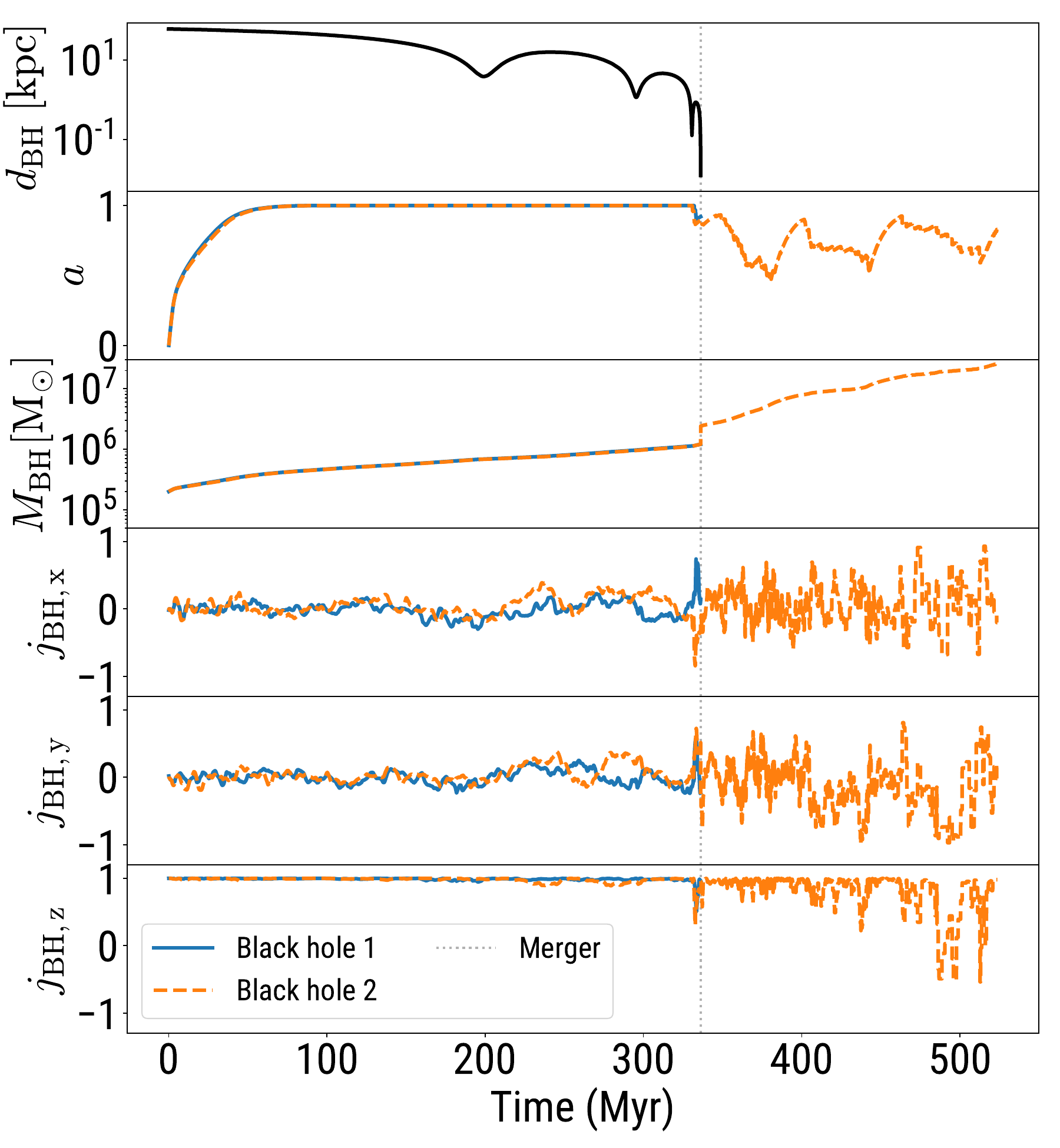}
    \includegraphics[width=0.48\textwidth, trim=0 0 0 0, clip]{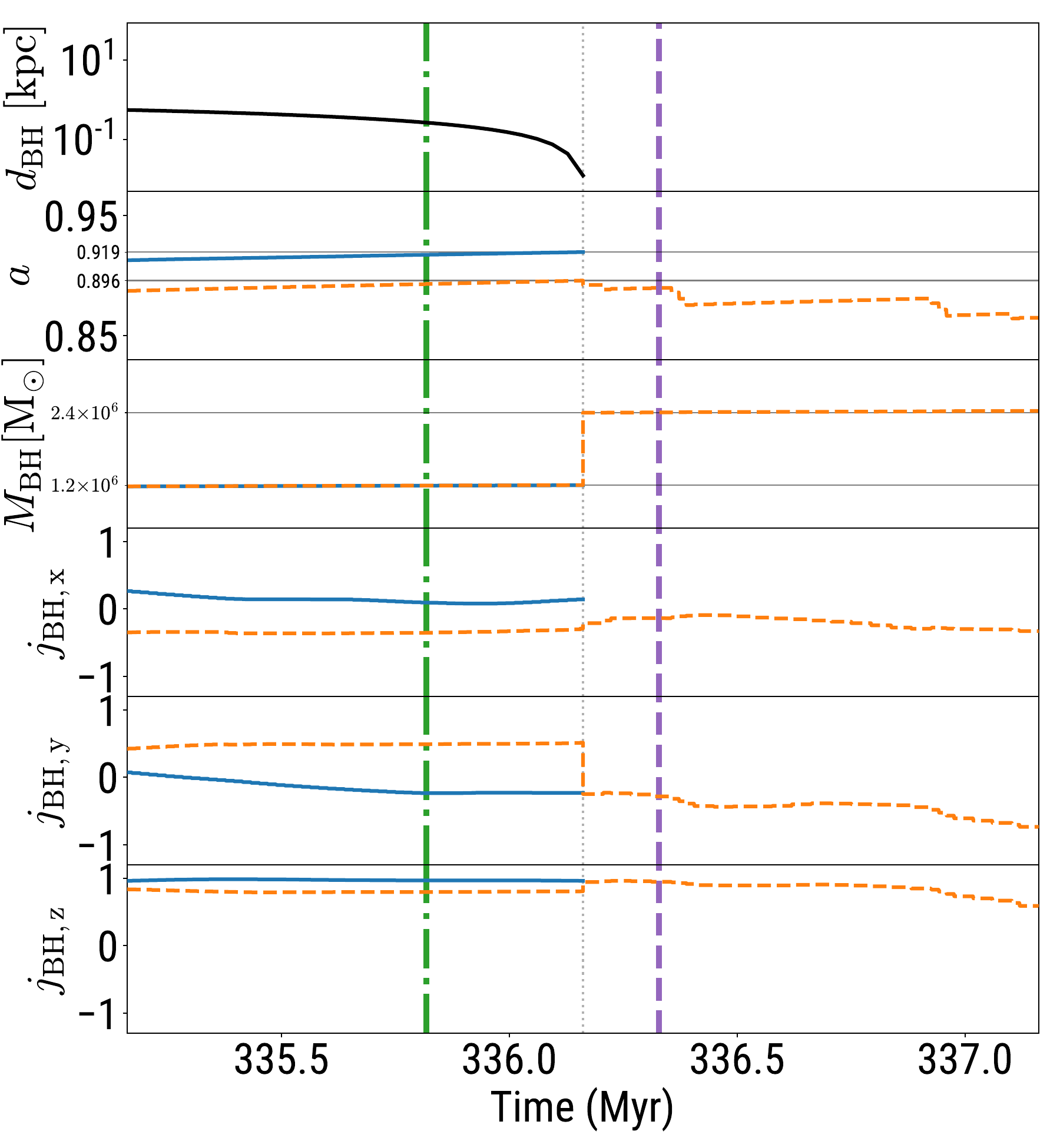}
    \caption{Time evolution of a few properties of the two BHs (each of them is depicted by either blue solid or orange dotted lines) in the idealised galaxy merger simulation. From top to bottom: separation distance between the two BHs; BH dimensionless spin parameter; BH mass; x, y, z component of the unit vector of the BH angular momentum. {\it Left panel}: entire simulated timespan. {\it Right panel}: timespan of 2 Myr, centered on the time of the merger (gray, dotted vertical line). The vertical green dash-dotted and purple dashed lines mark the instants in time illustrated in the top and bottom panel of Fig.~\ref{fig:snap_merger}, respectively.}
    \label{fig:merger_spin_details}
\end{figure*}

Fig.~\ref{fig:merger_spin_details} presents the evolution of the spins of the two BHs (one identified with the solid blue lines and the other with the dashed orange lines) in the idealised merger simulation. The left panel shows the evolution of quantities for the entire simulated timespan, whereas the right panel shows a 2 Myr window, centred on the merger instant, marked in both panels by the vertical grey dotted line. 

The first row from the top of Fig.~\ref{fig:merger_spin_details} shows the separation distance between the BHs. The galaxies undergo 3 pericenter passages before the BHs merge (left panel), at $\sim 336.15$ Myr (right panel).

The second row of Fig.~\ref{fig:merger_spin_details} follows the evolution of the spin parameter $a$. The left panel shows that both BHs follow an identical path from zero spin to maximal within 50 Myr (the two galaxies of the pair are identical). They are kept at maximal spin by accretion occurring consistently on the same plane, until the third pericenter passage. At this stage accretion becomes less coherent -- i.e. the direction of the local \am{} has larger variability -- and $a$ decreases slightly due to counter-rotating accretion conditions. In the right panel, second row, just before the merger instant we identify two progenitors with $a\sim 0.9$. The final remnant has $a\sim 0.9$.

The third row of Fig.~\ref{fig:merger_spin_details} shows the evolution of $\minsub{M}{BH}$. On the left, we see that the mass increases steadily and identically for the two BHs. The right panel allows us to identify two equal-mass progenitors with $\minsub{M}{BH}=1.2\times 10^6\msun$ merging into a remnant of $\minsub{M}{BH}=2.4\times10^6\msun$.

Rows 4 to 6 of Fig.~\ref{fig:merger_spin_details} show the three components of the BH spin direction. In the left panel we see that the two BHs consistently accrete from the same plane, hence their spins are kept very well aligned with the \am{} direction of the host galaxies, which is along $\minsub{\bm{j}}{z}$. The pericenter passages create some disturbance in the gas close to the BH, temporarily inducing the BH spin to tilt. This effect is more prominent close to the merger. In the right panel, we see that the BH spin components immediately before the merger are close to $\minsub{\bm{j}}{z}$, although with some disturbances induced by the local environment. Furthermore, the spin of the remnant is also aligned with $\minsub{\bm{j}}{z}$.

\begin{figure}
    \centering
    \includegraphics[width=0.48\textwidth, trim=0 0 0 0, clip]{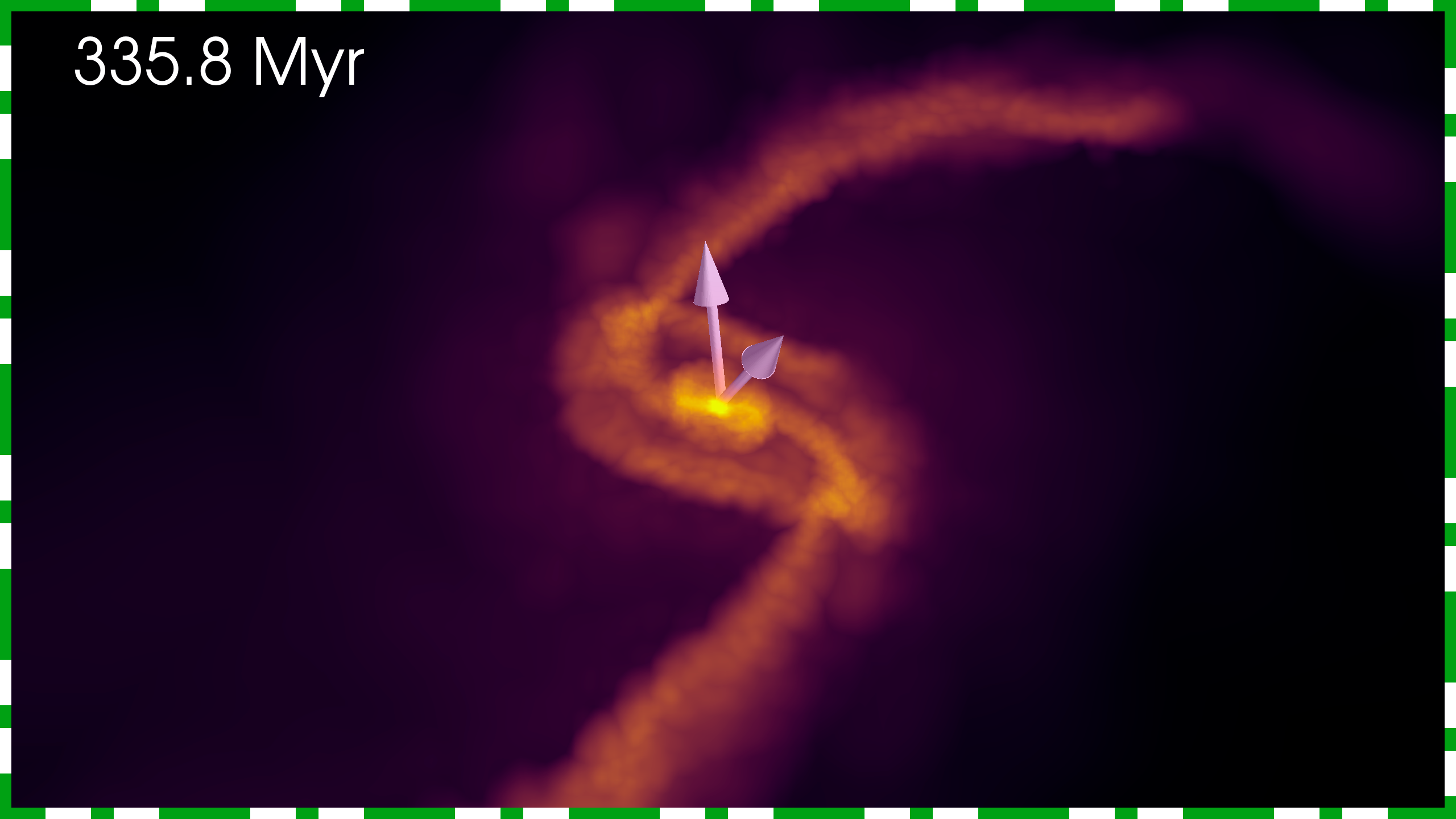}\hfill
    \includegraphics[width=0.48\textwidth, trim=0 0 0 0, clip]{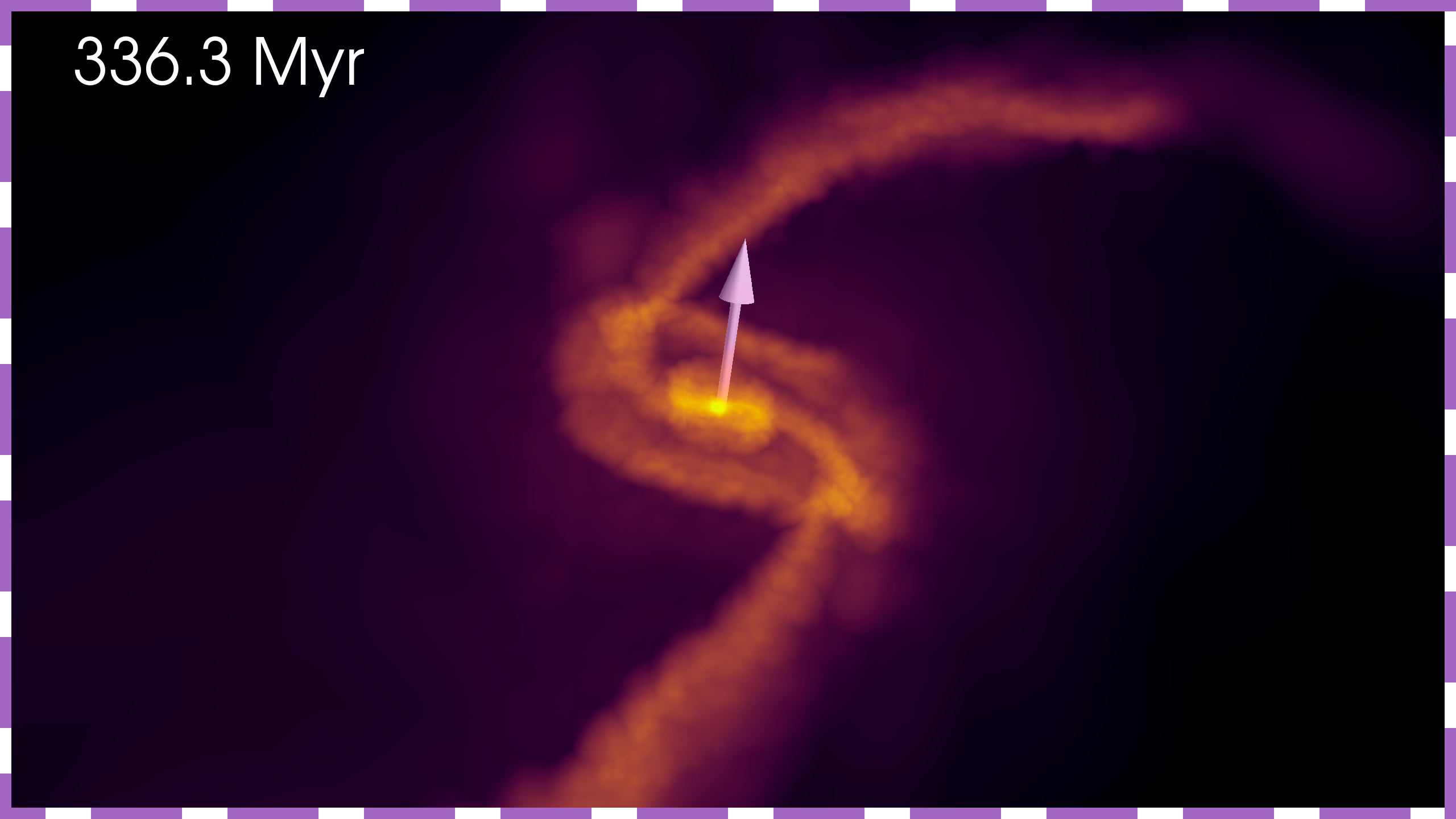}
    \caption{3D perspective volumetric rendering of the gas density in the idealised galaxy merger simulation. The field of view depicted in the figure spans a width of 55 kpc and a height of 31 kpc, as measured along a plane that intersects the centre of the rendering volume. {\it Top panel}: last snapshot before the merger. {\it Bottom panel}: first snapshot after merger. The arrows mark the instantaneous direction of the BH spin vector. (Movie online)}
    \label{fig:snap_merger}
\end{figure}

Fig.~\ref{fig:snap_merger} represents visually the process just described, using a volume rendering of the gas density in the simulation, for two simulation snapshots. The field of view depicted in the figure spans a width of 55 kpc and a height of 31 kpc, as measured along a plane that intersects the centre of the rendering volume, offering a perspective view of the merging galaxies. The arrows represent the instantaneous directions of the BH spins. The top panel of Fig.~\ref{fig:snap_merger} corresponds to the instant marked by the green dash-dotted line in the right panel of Fig.~\ref{fig:merger_spin_details}, showing the pre-merger configuration. The bottom panel of Fig.~\ref{fig:snap_merger} corresponds to the time marked by the purple dashed line, displaying the post-merger configuration.

The idealised galaxy merger test enables a simplified computation of the expected BH spin of the merger remnant to validate the result obtained. The pre-merger configuration (as shown in the right panels of Fig.~\ref{fig:merger_spin_details} and in the top panel of Fig.~\ref{fig:snap_merger}) can be approximated to the idealised configuration of an equal-mass BH binary with aligned spins treated in \cite{Rezzolla+2008a}. In this configuration, the final spin magnitude $\minsub{a}{f}$ depends only on the initial spin magnitudes $\minsub{a}{1}$ and $\minsub{a}{2}$ (and the final direction is equal to the initial one). Two BHs with $a\sim 0.9$ and spins in the same direction are indeed expected to end up in a merger remnant with $a\sim 0.9$ and spin along the original, common rotation axis, as observed in our test.

\subsection{Zoom-in simulations}\label{sec:zoomin_results}

\begin{figure}
    \centering
    \includegraphics[width=0.4\textwidth, trim=0 0 0 0, clip]{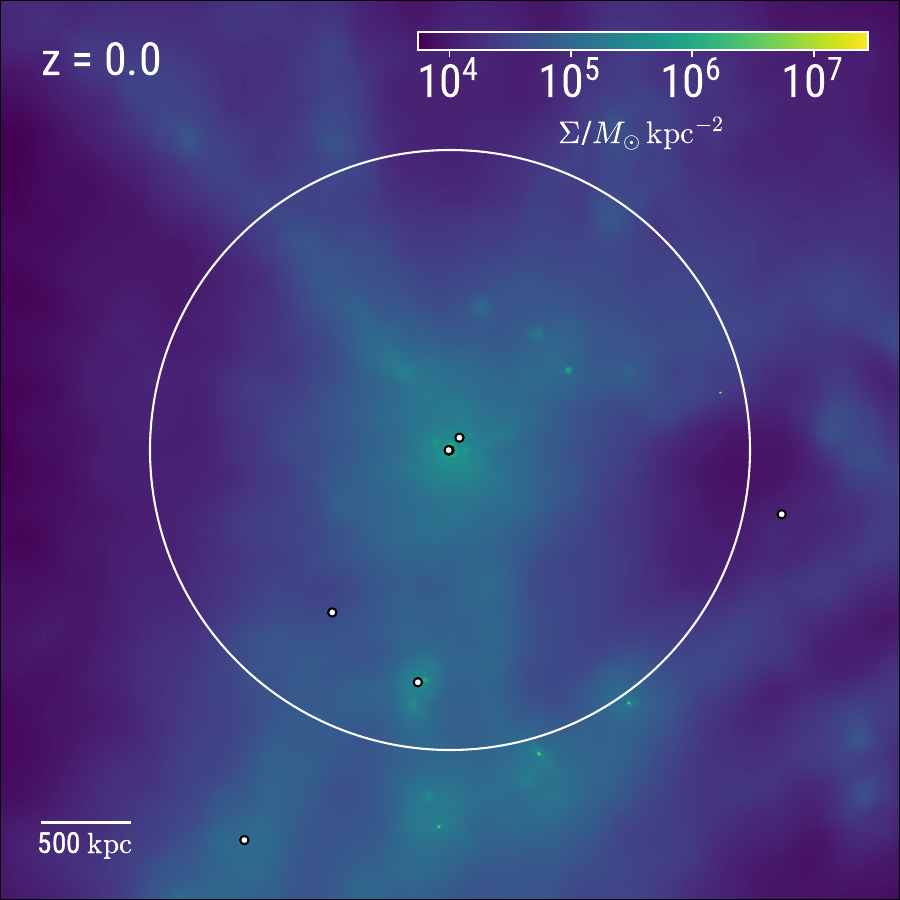}\hfill
    \includegraphics[width=0.4\textwidth, trim=0 0 0 0, clip]{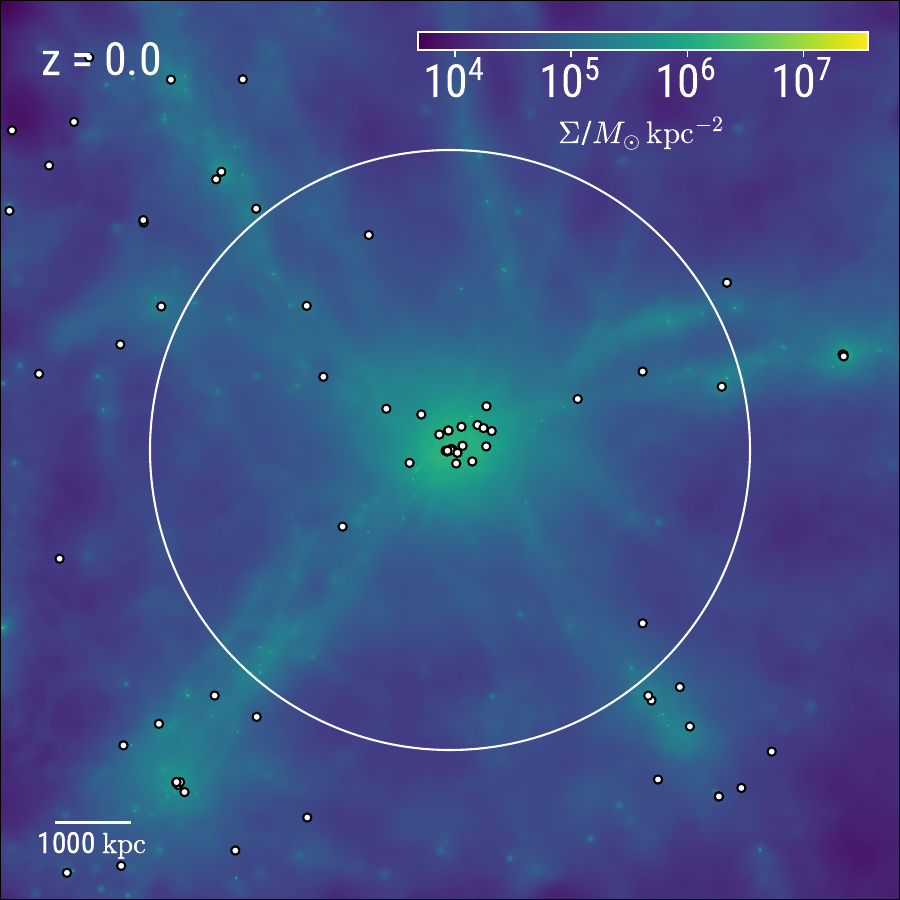}
    \caption{Gas surface density maps at $z=0$ for the \textsc{asin} (top) and \textsc{dfrogin} (bottom) runs. The white circle indicates a spherical region of radius $5\minsub{R}{200}$, centred on the target halo of the zoom-in region (i.e. the most massive at $z=0$). The white dots mark the positions of the BHs in the simulation.}
    \label{fig:filter_map}
\end{figure}
Fig.~\ref{fig:filter_map} presents projected gas density maps at $z=0$ of \textsc{asin} (top) and \textsc{dfrogin} (bottom). The white dots mark the positions of the BHs in the simulation. In what follows, we consider only the BHs that lie within a spherical region of radius $5\minsub{R}{200}$ at $z=0$, centred on the target halo of interest of the zoom-in region (i.e. the most massive one at $z=0$). $\minsub{R}{200}$ is the radius of the spherical volume, centred on the subhalo, whose average density is 200 times the critical density of the Universe. The region is marked by the white circle in Fig.~\ref{fig:filter_map}. The main purpose of this selection is to exclude BHs that are close the border of the high-resolution region.

\begin{figure}
    \centering
    \includegraphics[width=0.5\textwidth, trim=0 0 0 0, clip]{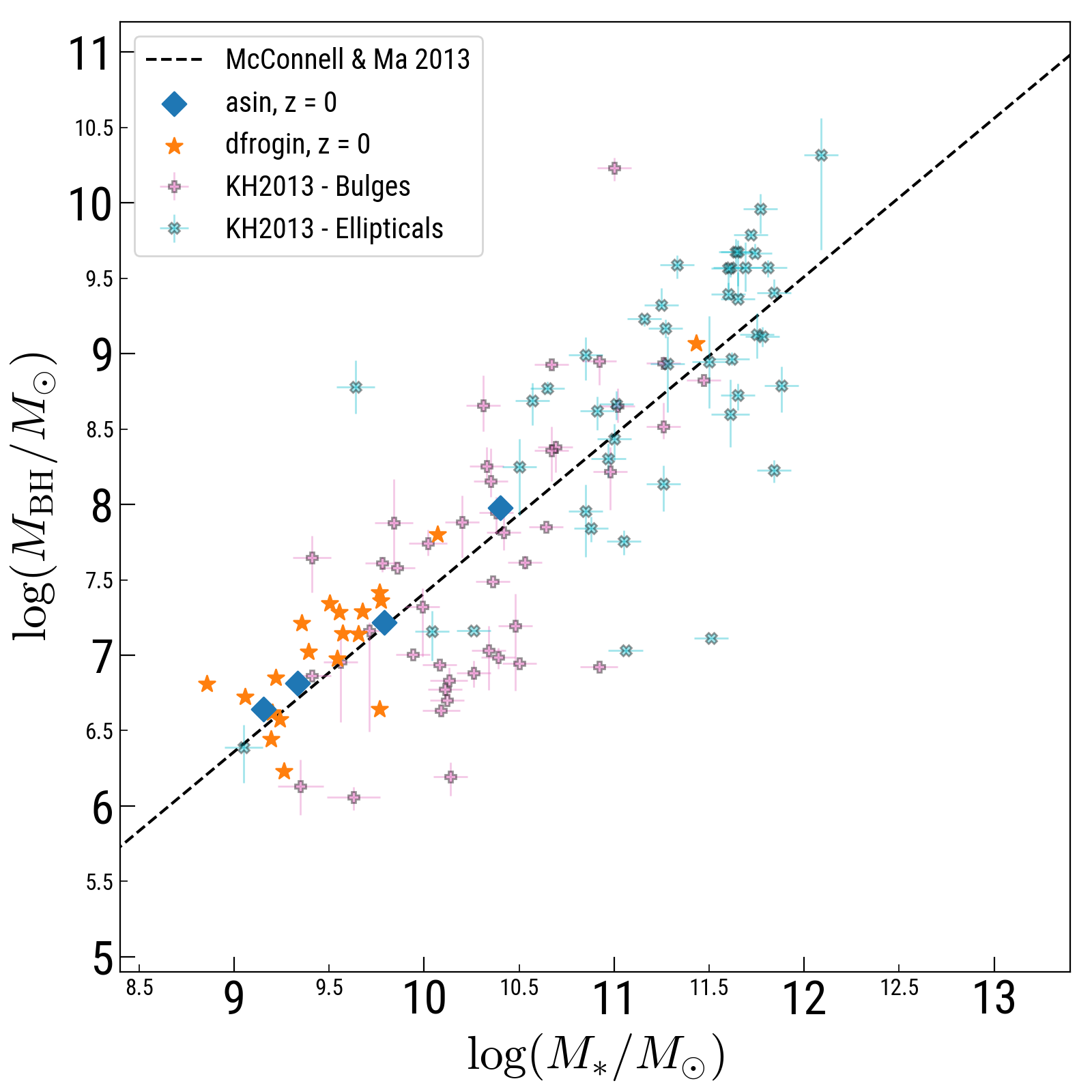}
    \caption{$M_{\rm BH}$ as a function of stellar mass $M_{\rm *}$, at $z=0$. Diamonds and stars represent BHs in \textsc{asin} and \textsc{dfrogin}, respectively. The dashed line shows the experimental fit by \citet{McConnell&Ma2013}, while crosses with associated uncertainties show data from \citet{Kormendy&Ho2013}.}
    \label{fig:magorrian_zoomins}
\end{figure}
Fig.~\ref{fig:magorrian_zoomins} shows the BH sample at $z=0$ in \textsc{asin} (diamonds) and \textsc{dfrogin} (stars) in the BH mass-stellar mass ($M_{\rm BH}-M_{\rm *}$) plane. The BHs are selected according to the criterion explained above; the sample is composed by 4 BHs in \textsc{asin} with $10^{6.5}\lesssim\minsub{M}{BH}/\msun\lesssim10^{8}$ and 21 BHs in \textsc{dfrogin}, with $10^{6}\lesssim\minsub{M}{BH}/\msun\lesssim10^{9}$. Each BH is associated to a subhalo identified with the sub-structure finder algorithm \textsc{SubFind}  \citep{Springel+2001, Dolag+2009} based on particle ID matching. Whenever more than one BH is associated to the same subhalo, the closest to the subhalo center is chosen. $M_{\rm *}$ is the stellar mass as computed by \textsc{SubFind}. The dashed line shows the experimental fit by \citet{McConnell&Ma2013}, while crosses with associated uncertainties show observations from \citet{Kormendy&Ho2013}.

\begin{figure}
    \centering
    \includegraphics[width=0.5\textwidth, trim=0 0 0 0, clip]{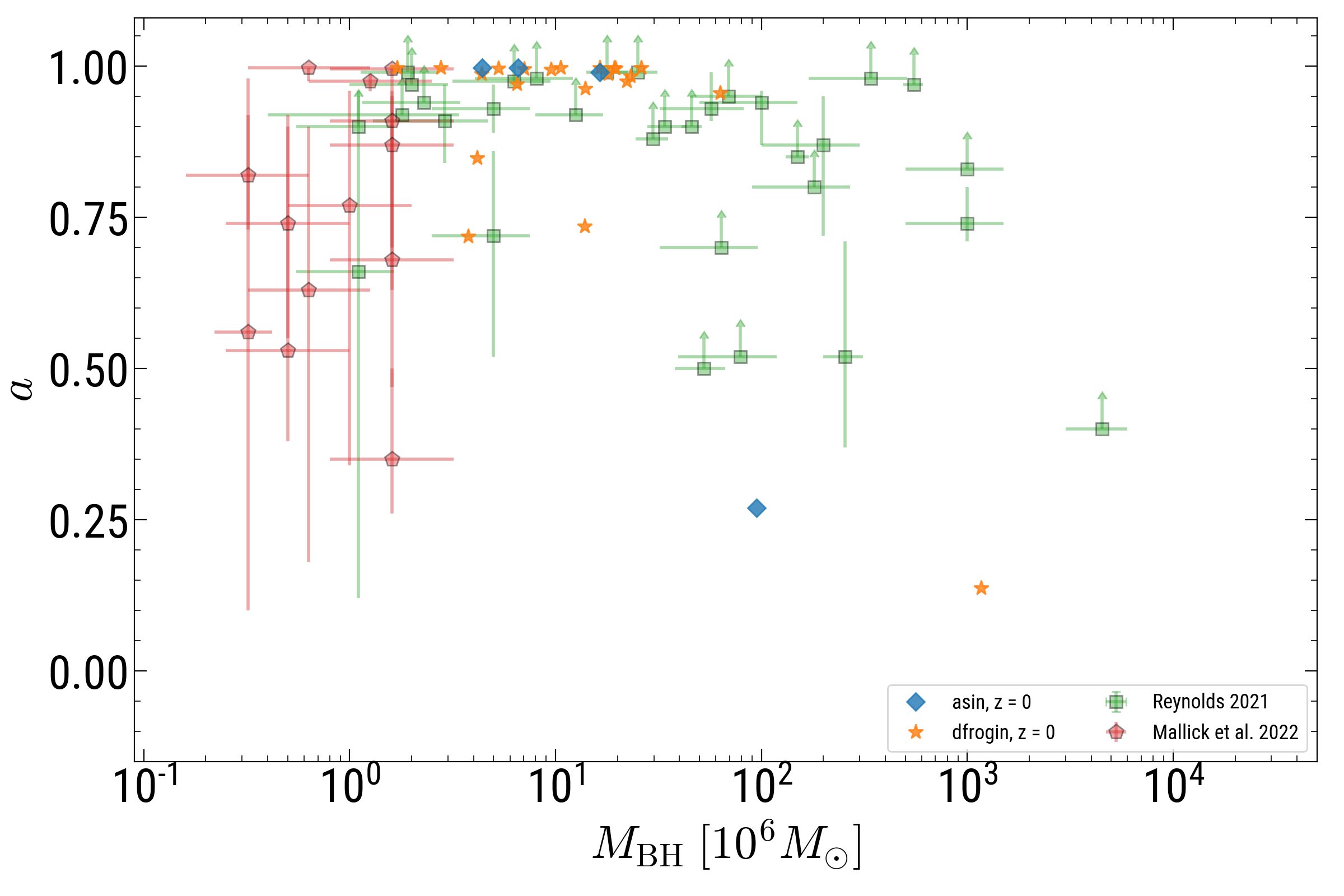}
    \caption{BH spin parameters $a$ of the selected BH sample in \textsc{asin} (diamonds) and \textsc{dfrogin} (stars) as a function of $\minsub{M}{BH}$, at $z=0$. The squares and pentagons show the collection of observational measurements of the BH spin parameter by \citet{Reynolds2021} and \citet{Mallick+2022}, respectively.}
    \label{fig:spin_statistics_zoomins}
\end{figure}
Fig.~\ref{fig:spin_statistics_zoomins} presents the spin parameters of the BHs in the simulated regions as a function of $\minsub{M}{BH}$ at $z=0$, with the same symbols used in Fig.~\ref{fig:magorrian_zoomins}. We compare our simulation output to the available observations that provide estimates of $a$ and BH mass \citep{Reynolds2021, Mallick+2022}, with associated uncertainties. We observe a range of BH masses ($10^{6}\lesssim\minsub{M}{BH}/\msun\lesssim 5\times10^{7}$) where BH spins tend to be larger than $a\sim0.70$, in both the simulated regions. Several of them are maximally spinning. For masses above $5\times 10^{7}\msun$, there are no maximally spinning BHs, whereas the most massive BH in each region systematically has a lower spin, $a\simeq0.25$ in \textsc{asin} and $a\simeq0.1$ in \textsc{dfrogin}.

\begin{figure*}
    \centering
    \includegraphics[width=1\textwidth, trim=0 0 0 0, clip]{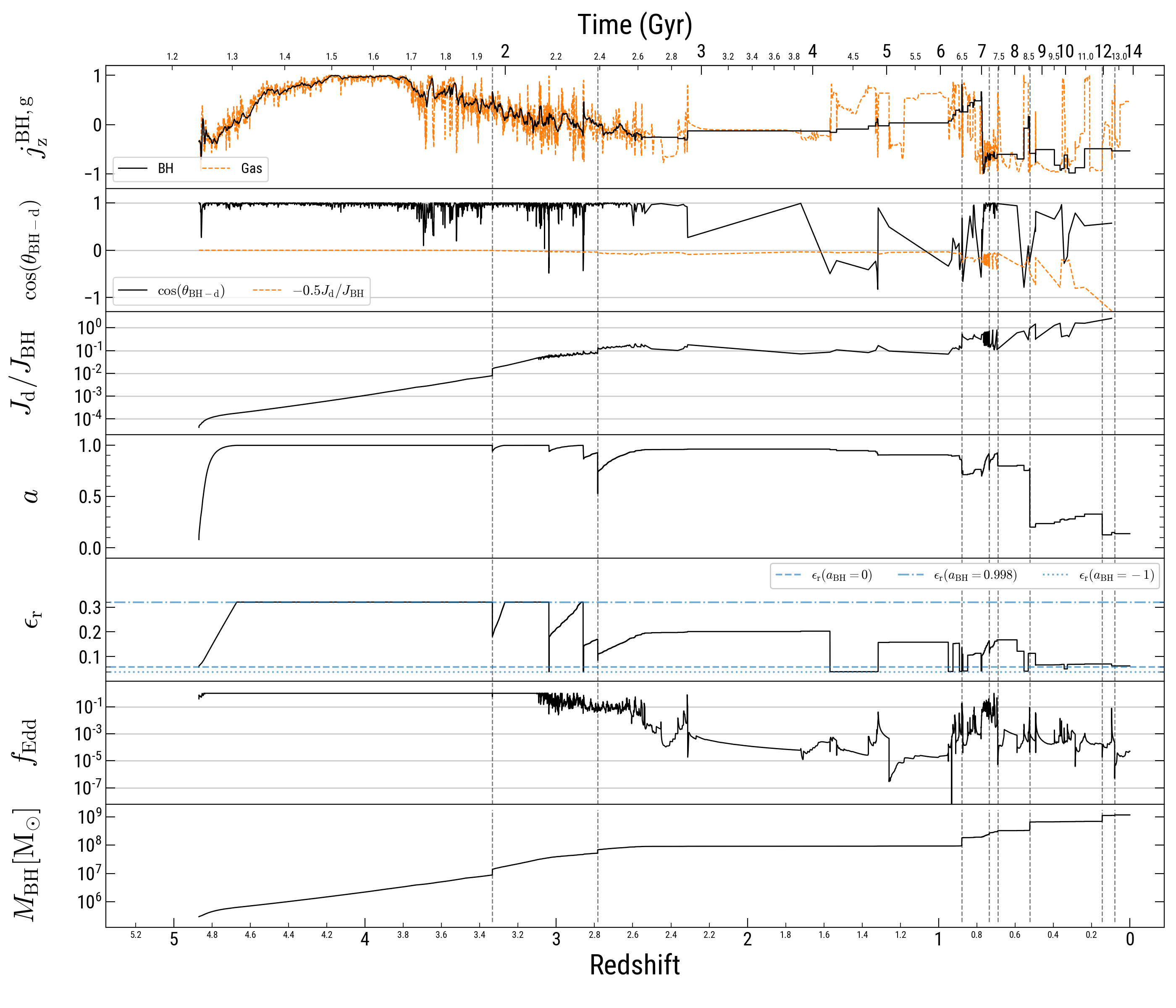}
    \caption{
    Summary of properties related to the most massive BH in the \textsc{dfrogin} run, as they evolve across time. 
    From top to bottom:
    z-component of the BH spin direction $\minsub{j}{BH,z}$ (black, solid line) and direction of the \am{} of the gas in the BH kernel $\minsub{j}{g,z}$ (orange, dashed line);
    cosine of the angle between the accretion disc and the BH spin at the beginning of each accretion episode $\cos{\theta_{\rm BH-d}}$ (i.e. left-hand side of Eq.~\eqref{eq:counter-align_condition}, black solid line) and $-J_{\rm d}/(2J_{\rm BH})$ (i.e. right-hand side of Eq.~\eqref{eq:counter-align_condition}, orange dashed line); 
    ratio of the magnitudes of the disc and BH angular momenta $\minsub{J}{d}/\minsub{J}{BH}$;
    BH dimensionless spin parameter $a$; 
    radiative efficiency of the accretion disc $\minsub{\epsilon}{r}$;    
    Eddington ratio $\minsub{f}{Edd}$;
    BH mass $\minsub{M}{BH}$. 
    }
    \label{fig:multiplot_asin}
\end{figure*}

Fig.~\ref{fig:multiplot_asin} shows the time evolution of a few BH properties predicted by the sub-resolution model. We restrict our analysis to the most massive BH of the sample in \textsc{dfrogin}, and focus on the relationship between BH spin and gas accretion as they evolve with time. 
The evolutionary tracks of the other BHs in the sample display features which are similar to those of our reference, most massive BH as for the interplay between the BH spin and the fuelling gas. The upper axis of Fig.~\ref{fig:multiplot_asin} marks the time since the Big Bang, while the lower axis shows the redshift $z$. Across the panels, the vertical dashed lines mark the occurrence of mergers. 

The top panel of Fig.~\ref{fig:multiplot_asin} shows the z-components of $\minsub{\bm{j}}{BH}$ (black, solid line) and $\minsub{\bm{j}}{g}$ (orange, dashed line). Before 1.7 Gyr, $\minsub{\bm{j}}{g}$ varies gradually with time, with a clear, average trend and limited scatter. From $\simeq1.7$ to $\simeq 2.6$ Gyr a trend in $\minsub{\bm{j}}{g}$ is still present but the scatter increases. $\minsub{\bm{j}}{BH}$ follows the average evolution of $\minsub{\bm{j}}{g}$. From $\simeq 2.6$ to $\simeq 6$ Gyr, $\minsub{\bm{j}}{g}$ exhibits drastic and sudden changes, whereas $\minsub{\bm{j}}{BH}$ remains stable with minimal variations. After $\simeq 6$ Gyr, $\minsub{\bm{j}}{BH}$ often undergoes abrupt variations due to the erratic behaviour of $\minsub{\bm{j}}{g}$.

The second row of Fig.~\ref{fig:multiplot_asin} shows $\cos \minsub{\theta}{BH-d}$ (black solid line), i.e. the left-hand-side term of Eq.~\eqref{eq:counter-align_condition}, as a function of time. The orange dashed line shows $-J_{\rm d}/(2J_{\rm BH})$, i.e. the right-hand side of Eq.~\eqref{eq:counter-align_condition}. Before $\simeq1.7$ Gyr each accretion episode is characterised by consistently small misalignment. Between $\simeq1.7$ and $\simeq2.6$ Gyr several accretion episodes display $\theta_{\rm BH-d}$ close to $\pi/2$. After $\simeq 2.6$ Gyr accretion episodes are mostly misaligned. This panel also allows us to easily identify counter-rotating accretion episodes, when $\cos{(\theta_{\rm BH-d})}<-J_{\rm d}/(2J_{\rm BH})$ (e.g. around 2.2 Gyr and 4.5 Gyr).

The third row of Fig.~\ref{fig:multiplot_asin} shows $J_{\rm d}/J_{\rm BH}$, the ratio between the accretion disc and the BH \am{} per accretion episode. This quantity controls how much a single accretion episode is able to modify the direction of the BH spin and ultimately determines to which degree the BH spin is able to follow the variations of $\minsub{\bm{j}}{g}$. Being the BH spin direction $\minsub{\bm{j}}{BH}^f$ after an accretion episode parallel to $\minsub{\bm{J}}{tot}=\minsub{\bm{J}}{d}+\minsub{\bm{J}}{BH}$, if $\minsub{J}{d}\ll\minsub{J}{BH}$, then $\minsub{\bm{j}}{BH}^{f}\sim\minsub{\bm{j}}{tot}\sim\minsub{\bm{j}}{BH}$ (i.e. the direction change is negligible; Sec.~\ref{sec:spin_udpate_iteration}). Conversely, if $\minsub{J}{d}\gg\minsub{J}{BH}$, then $\minsub{\bm{j}}{BH}^{f}\sim\minsub{\bm{j}}{d}$, i.e. $\minsub{\bm{j}}{BH}$ aligns with the direction imparted by $\minsub{\bm{j}}{d}$ (and hence $\minsub{\bm{j}}{g}$). In an intermediate configuration $\minsub{\bm{j}}{BH}$ only partially aligns with $\minsub{\bm{j}}{d}$. 
The latter situation is expected to occur before $\simeq6$ Gyr, when $\minsub{J}{d}/\minsub{J}{BH}\lesssim 0.1$. 
However, at $t\gtrsim 6$ Gyr $\minsub{J}{d}/\minsub{J}{BH}$ sometimes approaches 1; as a result, $\minsub{\bm{j}}{BH}$ exhibits larger variations as a response to large changes in $\minsub{\bm{j}}{g}$. Although here we inspect the evolution of $\minsub{J}{d}/\minsub{J}{BH}$ and ${\theta_{\rm BH-d}}$ for one BH as an example, we properly quantify how the change $\minsub{\bm{j}}{BH}$ depends on these two quantities by analysing accretion episodes statistically in the cosmological box (Sec.~\ref{sec:cosmo_box_results}).

The fourth row of Fig.~\ref{fig:multiplot_asin} shows the BH dimensionless spin parameter $a$. Its evolution is characterised by a maximally-spinning phase until $t\simeq 1.95$ Gyr ($z\gtrsim 3.3$). The largest spin-downs occur because of mergers, although we also observe counter-rotating phases that decrease the spin due to gas accretion (e.g. at $t \simeq 2.2$ and $\simeq2.35$ Gyr, as well as around $4.5$ Gyr).

The fifth row of Fig.~\ref{fig:multiplot_asin} plots the radiative efficiency, which depends on $a$ (see Eq.~\ref{eq:radiative_efficiency}). The dash-dotted, dashed and dotted lines mark the values of efficiency for $a=0.998$, 0 and for a counter-rotating episode on a BH with $a=1$, respectively. The maximally-spinning period before $z\sim 3.5$ corresponds to a maximal efficiency of 0.32, whereas later times are characterised by a lower efficiency. Counter-rotating accretion conditions are clearly visible as a drastic decrease in efficiency, close to the minimum theoretical value marked by the dotted line. 

The sixth row of Fig.~\ref{fig:multiplot_asin} illustrates the Eddington ratio $\minsub{f}{Edd}$. Accretion is Eddington-limited soon after the BH is seeded ($z \sim 4.9$). $\minsub{f}{Edd}$ is typically $\gtrsim10^{-1}$ until $z\sim 2.6$. Between $0.9\lesssim z\lesssim 2.6$, $\minsub{f}{Edd}\ll10^{-1}$ for most of the time. A few highly accreting ($\minsub{f}{Edd}\simeq10^{-1}$) episodes occur in proximity of the mergers at $z \sim 0.9$ and $z \sim 0.7$, and lead to significant reorientation of the BH spin (see first row of Fig.~\ref{fig:multiplot_asin}). After $z\sim 0.7$ $\minsub{f}{Edd}$ is mostly $\lesssim10^{-3}$. Comparing with the evolution of $a$ (fourth row of Fig.~\ref{fig:multiplot_asin}), we observe that the largest $\minsub{f}{Edd}$, the largest is the change induced in $a$.

The last row of Fig.~\ref{fig:multiplot_asin} plots $\minsub{M}{BH}$. The BH increases its mass by three orders of magnitude during the highly-accreting phase at $z\gtrsim 2.6$. After $z\sim 2.6$, the BH gets more massive (by about one order of magnitude) mostly due to mergers. 

\subsection{Cosmological box}\label{sec:cosmo_box_results}

For the analysis of \textsc{Box4} we consider all the BHs in the cosmological volume and we select them at $z=0$. We associate each BH to a subhalo  based on particle ID matching using \textsc{SubFind} and whenever more than one BH is associated to the same subhalo, the closest to the subhalo center is chosen. The sample includes 1790 BHs, in a mass range between $6\times10^6$ and $2\times10^{10}\msun$.

\begin{figure}
    \centering
    \includegraphics[width=0.5\textwidth, trim=0 0 0 0, clip]{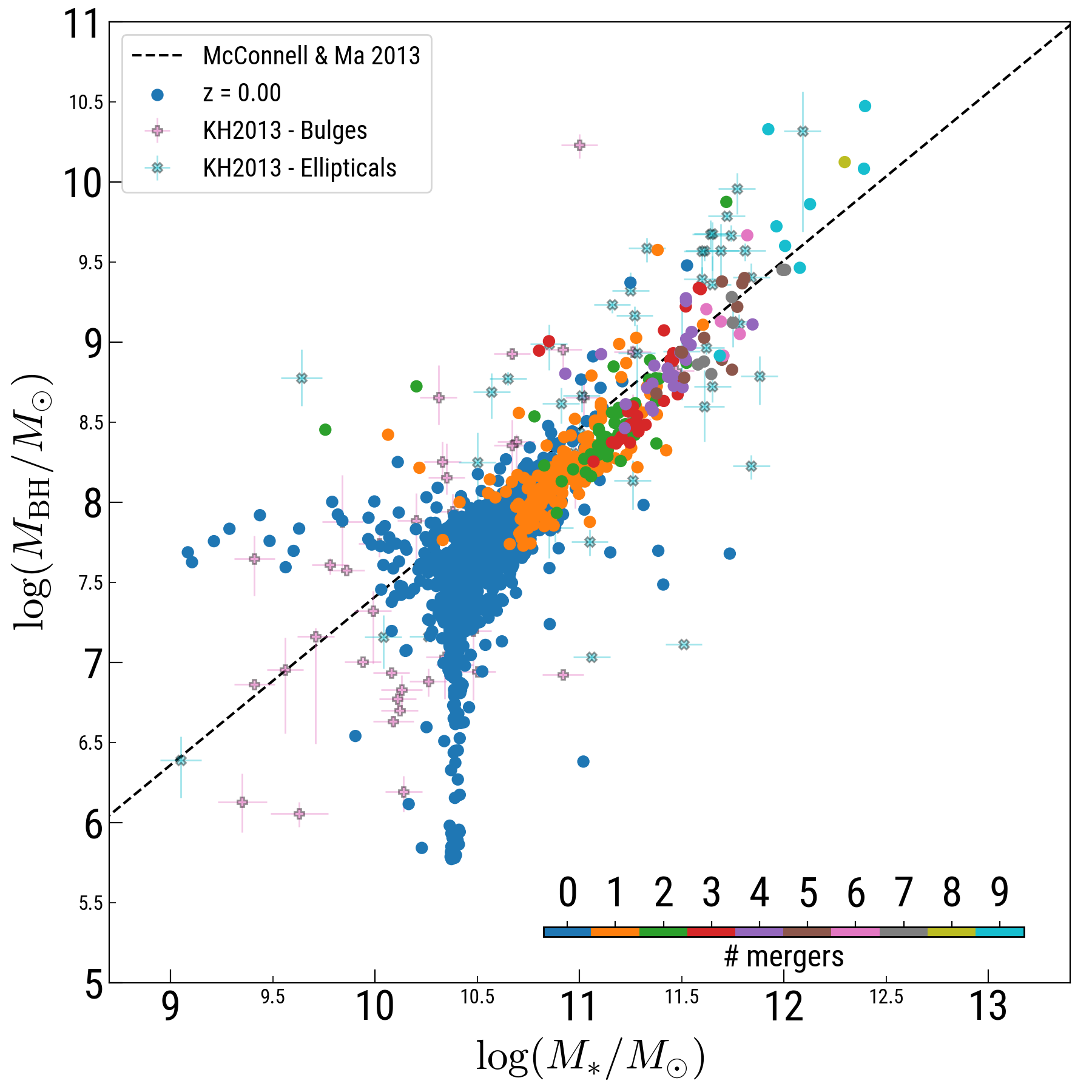}
    \caption{$M_{\rm BH}$ as a function of stellar mass $M_{\rm *}$ for the \textsc{Box4} run. Circles represent the simulated sample at $z=0$, while the observational points are as in Fig.~\ref{fig:magorrian_zoomins}. The points are colour-coded according to the number of mergers BH have undergone.}
    \label{fig:magorrian_box}
\end{figure}
Fig.~\ref{fig:magorrian_box} frames the \textsc{Box4} BH sample at $z\sim 0$ in the $\minsub{M}{BH}-\minsub{M}{*}$ plane. $\minsub{M}{*}$ corresponds to the subhalo stellar mass as computed by \textsc{SubFind}. We compare our sample to observations by \cite{Reynolds2021} and \cite{Mallick+2022} (as in Fig.~\ref{fig:magorrian_zoomins}). Each point is colour-coded according to the number of mergers of the BH (including their progenitors). The number increases with increasing BH mass.

\begin{figure*}
    \centering
    \includegraphics[width=\textwidth, trim=0 0 0 0, clip]{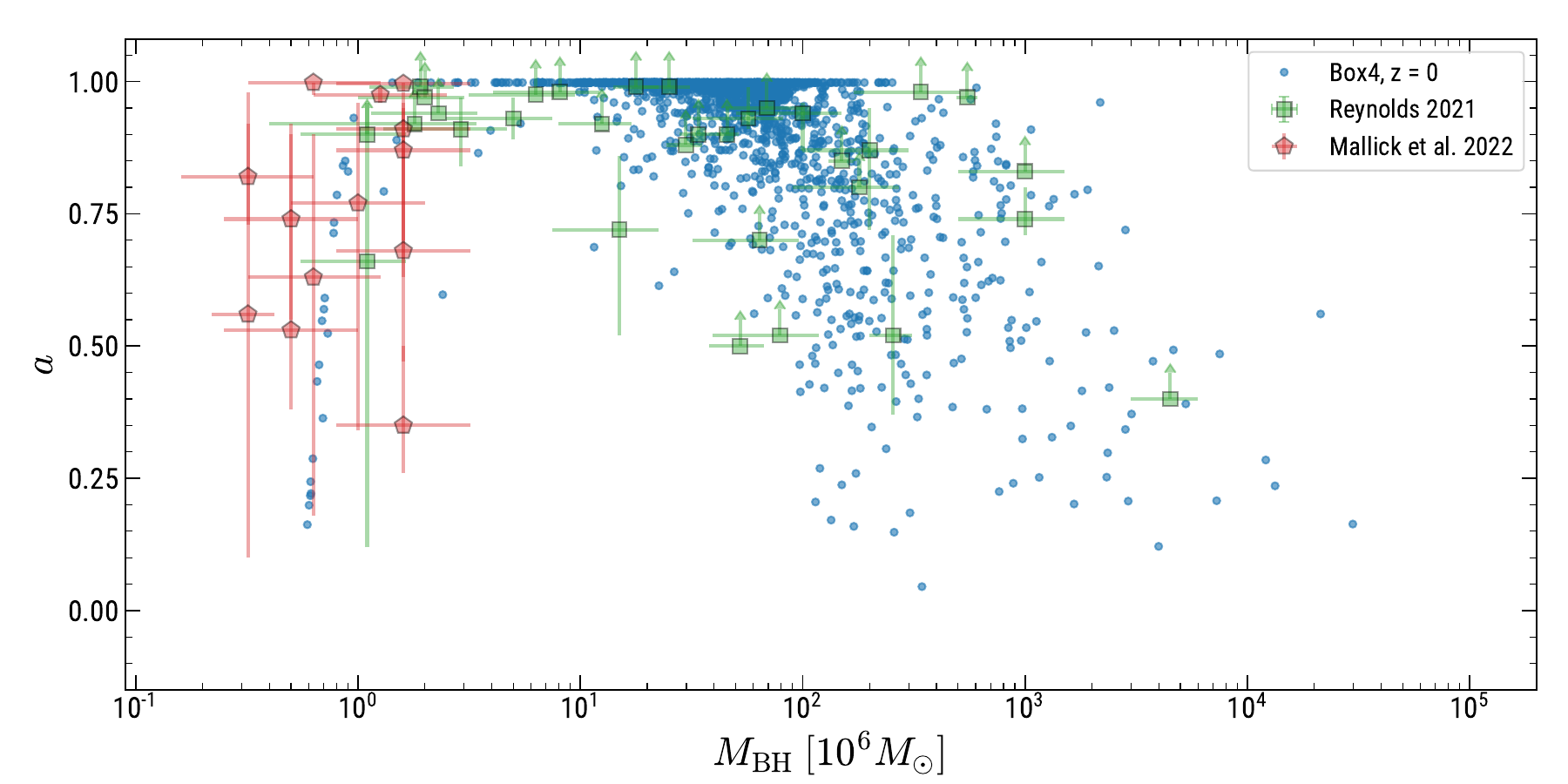}
    \caption{Distribution of BH spin parameters as a function of the BH mass in \textsc{Box4} (circles), at $z=0$. The squares and pentagons show the collection of observational measurements of the BH spin parameter by \citet{Reynolds2021} and \citet{Mallick+2022}, respectively.}
    \label{fig:spin_statistics_box}
\end{figure*}
In Fig.~\ref{fig:spin_statistics_box} we show how BH spin parameters change as a function of $M_{\rm BH}$ in \textsc{Box4} (similar to Fig.~\ref{fig:spin_statistics_zoomins}). The simulated BHs are indicated by circles. We compare our simulation output to the observational data by \cite{Reynolds2021} and \cite{Mallick+2022}. The BH sample is much more numerous than in the zoom-in regions and the mass range extends further on the high-end, up to $2\sim 10^{10}\msun$. We identify three mass ranges in which we observe different distributions of $a$. Close to the seeding mass ($\sim 5.5 \times 10^{5}\msun$), we can see a steep increase of $a$ with $\minsub{M}{BH}$. The BH population characterised by $10^{6}\lesssim\minsub{M}{BH}/\msun\lesssim 2\times10^{7}$ shows a systematic tendency for highly spinning BHs ($a\gtrsim0.85$), with most of them close to the maximal value. BHs with masses above $2\times10^{7} \msun$ display a wider range of $a$, extending as low as $a\sim0.1$. The robustness of this result is consolidated by the larger number of BHs probing this high-mass regime with respect to the zoom-in simulations. We also note that for $\minsub{M}{BH}\gtrsim 5\times 10^{8} \msun$ there are no maximally spinning BHs. 

\begin{figure*}
    \centering
    \includegraphics[width=\textwidth, trim=0 0 0 0, clip]{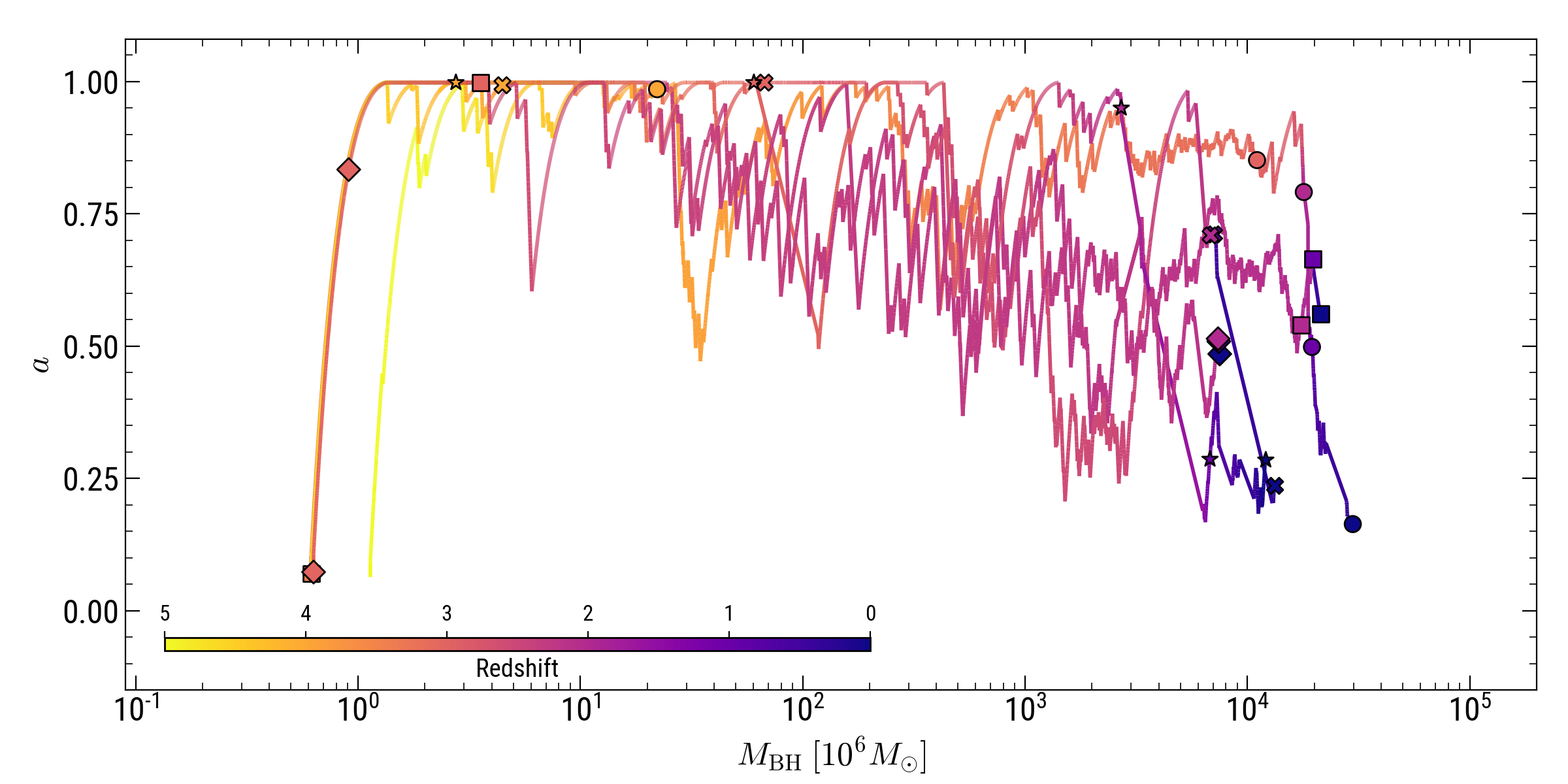}
    \caption{BH dimensionless spin parameter $a$ as a function of $\minsub{M}{BH}$, for the five most massive BHs at $z=0$, for the \textsc{Box4} run. Each line corresponds to one BH and is colour-coded by redshift. The symbols mark the position of each BH in the plot at a few specific instants in time, corresponding to $z={0,1,2,3,4}$, colour-coded accordingly.}
    \label{fig:10mostmassive_box}
\end{figure*}
Fig.~\ref{fig:10mostmassive_box} shows the evolution of $a$ of the five most massive BHs in the sample as a function of mass. The redshift is encoded in the colour gradient of each curve. The symbols mark the position of each BH in the plot at a few specific instants in time, corresponding to $z={0,1,2,3,4}$, colour-coded accordingly.
As soon as the BHs are seeded, they undergo a phase of rapid increase of $a$, then they reach and maintain a maximally spinning state. We also notice that when $\minsub{M}{BH}$ is between $10^{6}$ and $2\times10^7\msun$, BHs undergo short transitory periods of spin-down due to counter-rotating accretion or mergers. On the other hand, $a$ reaches again the maximal state afterwards, highlighting that co-rotating accretion is dominant. After the BHs overcome the $2\times10^{7}\msun$ mass scale, $a$ exhibits a tendency to decrease. In addition, we observe large and sudden changes due to mergers (when also the mass increases significantly at the same time). A binary with BH spins oriented towards opposite directions results in a severe spin-down of the remnant compared to the state immediately before the merger. The wider distribution of $a$ at the highest masses in Fig.~\ref{fig:spin_statistics_box} suggests that spin-down due to counter-rotating accretion and mergers occurs frequently.

\begin{figure*}
    \centering
    \includegraphics[width=1\textwidth, trim=30 0 90 0, clip]{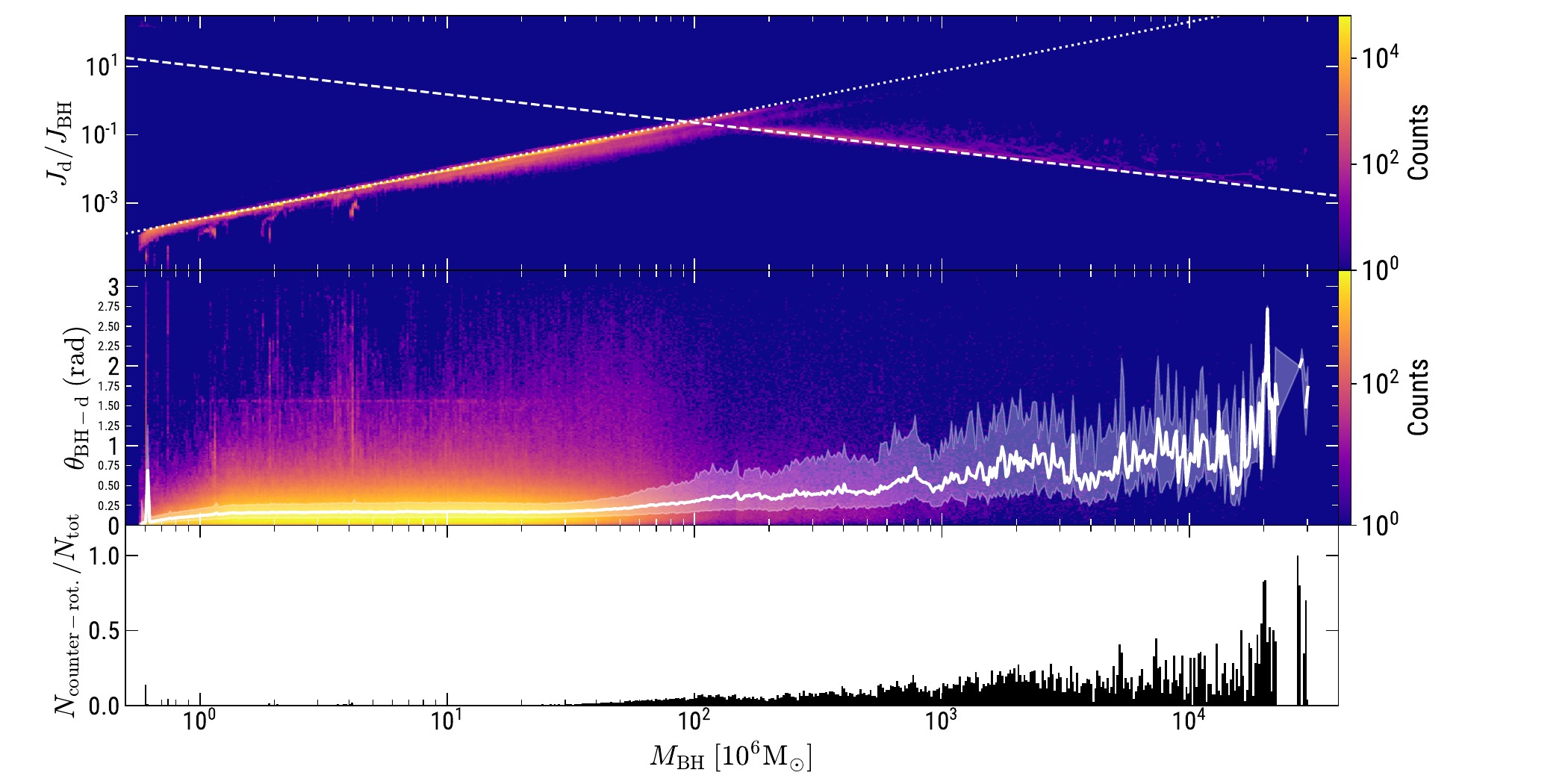}
    \caption{Statistical analysis of a few key properties of the accretion episodes occurred in the \textsc{Box4} run. The entire set of accretion episodes occurred during the simulation has been considered (i.e. for every BH and at every redshift). The top (middle) panel shows a 2D histogram of the values of $\minsub{J}{d}/\minsub{J}{BH}$ ($\minsub{\theta}{BH-d}$) as a function of mass. In the top panel, the dashed line pinpoints $\minsub{J}{d}/\minsub{J}{BH}\propto\minsub{M}{BH}^{-37/45}$ (for the self-gravitating case, Eq.~\ref{eq:JdJBHratio_sg_numeric}); the dotted line shows $\minsub{J}{d}/\minsub{J}{BH}\propto\minsub{M}{BH}^{23/16}$ (non self-gravitating case, Eq.~\ref{eq:JdJBHratio_numeric}). $a=0.998$ and $\minsub{f}{Edd}=1$ are assumed to plot these reference lines. The bottom panel shows the fraction of counter-rotating accretion episodes over the total, per BH mass bin.}
    \label{fig:binned_disc_prop}
\end{figure*}
In Fig.~\ref{fig:binned_disc_prop} we carry out a statistical analysis of a few key properties of the accretion episodes occurred in the \textsc{Box4} run, to gain insight on the mechanisms with which gas accretion drives the trends observed in Fig.~\ref{fig:spin_statistics_box} and \ref{fig:10mostmassive_box}. For the analysis, the entire set of accretion episodes occurred during the simulation is considered (i.e. for every BH at every redshift). The top panel of Fig.~\ref{fig:binned_disc_prop} shows a 2D histogram where accretion episodes are binned according to their values of $\minsub{J}{d}/\minsub{J}{BH}$ and $\minsub{M}{BH}$. Each 2D bin is colour-coded by the number of accretion episodes in that bin. The dashed and dotted lines represent the dependence of $\minsub{J}{d}/\minsub{J}{BH}$ on mass from Eq.~\ref{eq:JdJBHratio} (i.e. $\propto\minsub{M}{BH}^{-37/45}$ for the self-gravitating case and $\propto\minsub{M}{BH}^{23/16}$ for the standard case), assuming $a=0.998$ and $\minsub{f}{Edd}=1$. The distribution observed in $\minsub{J}{d}/\minsub{J}{BH}$ at fixed mass bin is due to different values of $a$ and $\minsub{f}{Edd}$. However, $\minsub{J}{d}/\minsub{J}{BH}$ depends weakly on the latter two quantities, while it depends on $\minsub{M}{BH}$ quite strongly (see Eq.~\ref{eq:JdJBHratio_numeric}). Above $\minsub{M}{BH}\sim10^8\msun$ accretion occurs mostly in the self-gravitating regime (i.e. $\minsub{R}{sg}<\minsub{R}{w}$).
The middle panel of Fig.~\ref{fig:binned_disc_prop} shows a 2D histogram where accretion episodes are binned according to their values of $\minsub{\theta}{BH-d}$ and $\minsub{M}{BH}$. The solid white line indicates the median value of $\minsub{\theta}{BH-d}$ per BH mass bin, whereas the shaded region represents the 25th and 75th percentile of the distribution of $\minsub{\theta}{BH-d}$ in that BH mass bin. We observe that accretion episodes are characterised predominantly by small misalignment (75\% of them has $\minsub{\theta}{BH-d}\lesssim0.25\text{ rad}\sim15^{\circ}$) below $\minsub{M}{BH}\sim10^8\msun$. Above this mass threshold, the distribution broadens with increasing mass, showing that large misalignment is increasingly more probable. However, we caution that in this regime less accretion episodes occur, therefore the significance of this trend is limited by low-number statistics. 
Note that when $\minsub{J}{d}\ll\minsub{J}{BH}$, the minimum angle required for an episode to satisfy condition \eqref{eq:counter-align_condition} is $\minsub{\theta}{BH-d}=\pi/2$. The minimum angle is larger for larger $\minsub{J}{d}/\minsub{J}{BH}$, whereas for $\minsub{J}{d}\geq2\minsub{J}{BH}$ accretion will be always co-rotating regardless the initial misalignment. Since a large misalignment is more probable at high BH masses (middle panel of Fig.~\ref{fig:binned_disc_prop}) whereas $\minsub{J}{d}/\minsub{J}{BH}$ decreases with mass above $\minsub{M}{BH}\sim10^{8}\msun$ (top panel of Fig.~\ref{fig:binned_disc_prop}), counter-rotating accretion episodes are more likely. 
In the bottom panel of Fig.~\ref{fig:binned_disc_prop} we plot the fraction of accretion episodes that are counter-rotating with respect to the total number per BH mass bin. This fraction increases with increasing mass and it is as high as 0.5 for the highest mass bins. 

It is now possible to assess the contribution of gas accretion to the trends discussed in Fig.~\ref{fig:spin_statistics_box} and \ref{fig:10mostmassive_box} in light of the results shown in Fig.~\ref{fig:binned_disc_prop}. Below $\minsub{M}{BH}\sim10^8\msun$, most of the accretion episodes are co-rotating, therefore spin-up is favoured. Counter-rotating accretion episodes do occur, but the decrease in spin is transitory and co-rotating accretion leads the spin back to maximal. Conversely, above $\minsub{M}{BH}\sim10^8\msun$ the probability of counter-rotating accretion (and hence spin-down) increases as a function of mass.

\begin{figure*}
    \centering
    \includegraphics[width=1\textwidth, trim=25 0 70 0, clip]{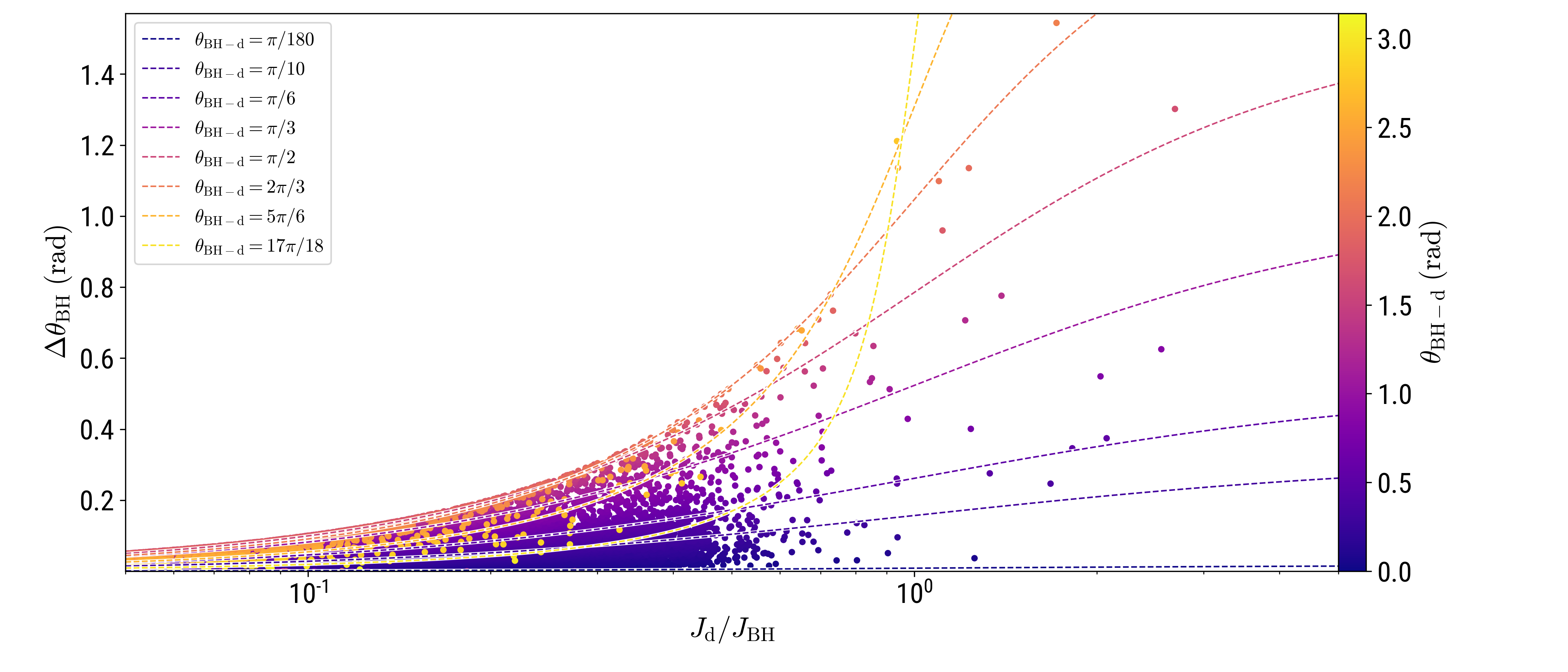}
    \caption{BH spin direction variation per accretion episode (i.e. $\Delta\minsub{\theta}{BH}=\minsub{\bm{j}}{BH}\cdot\minsub{\bm{j}}{BH}^{f}$) as a function of $\minsub{J}{d}/\minsub{J}{BH}$, colour-coded by $\minsub{\theta}{BH-d}$, the misalignment angle between disc and BH angular momenta at the beginning of the episode. The dashed lines illustrate the analytical dependence of $\minsub{\bm{j}}{BH}\cdot\minsub{\bm{j}}{BH}^{f}$ on $\minsub{J}{d}/\minsub{J}{BH}$, computed using Eq.~\ref{eq:Jtot}.}
    \label{fig:spin_tilt}
\end{figure*}
In Fig.~\ref{fig:spin_tilt} we quantify the direction variation imparted to the BH spin as a function of the properties of each accretion episode. The plot shows the angle $\Delta\minsub{\theta}{BH}$ between the direction of the BH spin before and after each accretion episode (i.e. $\minsub{\bm{j}}{BH}\cdot\minsub{\bm{j}}{BH}^{f}$) in the \textsc{Box4} run, as a function of $\minsub{J}{d}/\minsub{J}{BH}$. The accretion episodes (circles) are colour-coded by $\minsub{\theta}{BH-d}$, i.e. the BH spin-disc misalignment at the beginning of the episode. The dashed lines indicate the analytical dependence computed using Eq.~\eqref{eq:Jtot} and expressing $\minsub{\bm{j}}{BH}\cdot\minsub{\bm{j}}{BH}^{f}$ as a function of $\minsub{J}{d}/\minsub{J}{BH}$, at fixed $\minsub{\theta}{BH-d}$. If $\minsub{J}{d}\ll\minsub{J}{BH}$ then $\Delta\minsub{\theta}{BH}\sim0$, regardless $\minsub{\theta}{BH-d}$. If $\minsub{J}{d}\gg\minsub{J}{BH}$, then $\Delta\minsub{\theta}{BH}\sim\minsub{\theta}{BH-d}$. Increasing values of $\minsub{J}{d}/\minsub{J}{BH}$ induce larger alignment of $\minsub{\bm{j}}{BH}$ with $\minsub{\bm{j}}{d}$ per accretion episode.
Fig.~\ref{fig:spin_tilt} also shows that, overall, $\minsub{J}{d}/\minsub{J}{BH}\gtrsim1$ is rare. Therefore complete alignment of $\minsub{\bm{j}}{BH}$ with $\minsub{\bm{j}}{d}$ ($\Delta\minsub{\theta}{BH}=\minsub{\theta}{BH-d}$) never occurs in a single accretion episode, i.e. accretion episodes lead at most to partial alignment with the instantaneous direction of the accreting gas.

\begin{figure*}
    \centering
    \includegraphics[width=\textwidth, trim=0 0 0 0, clip]{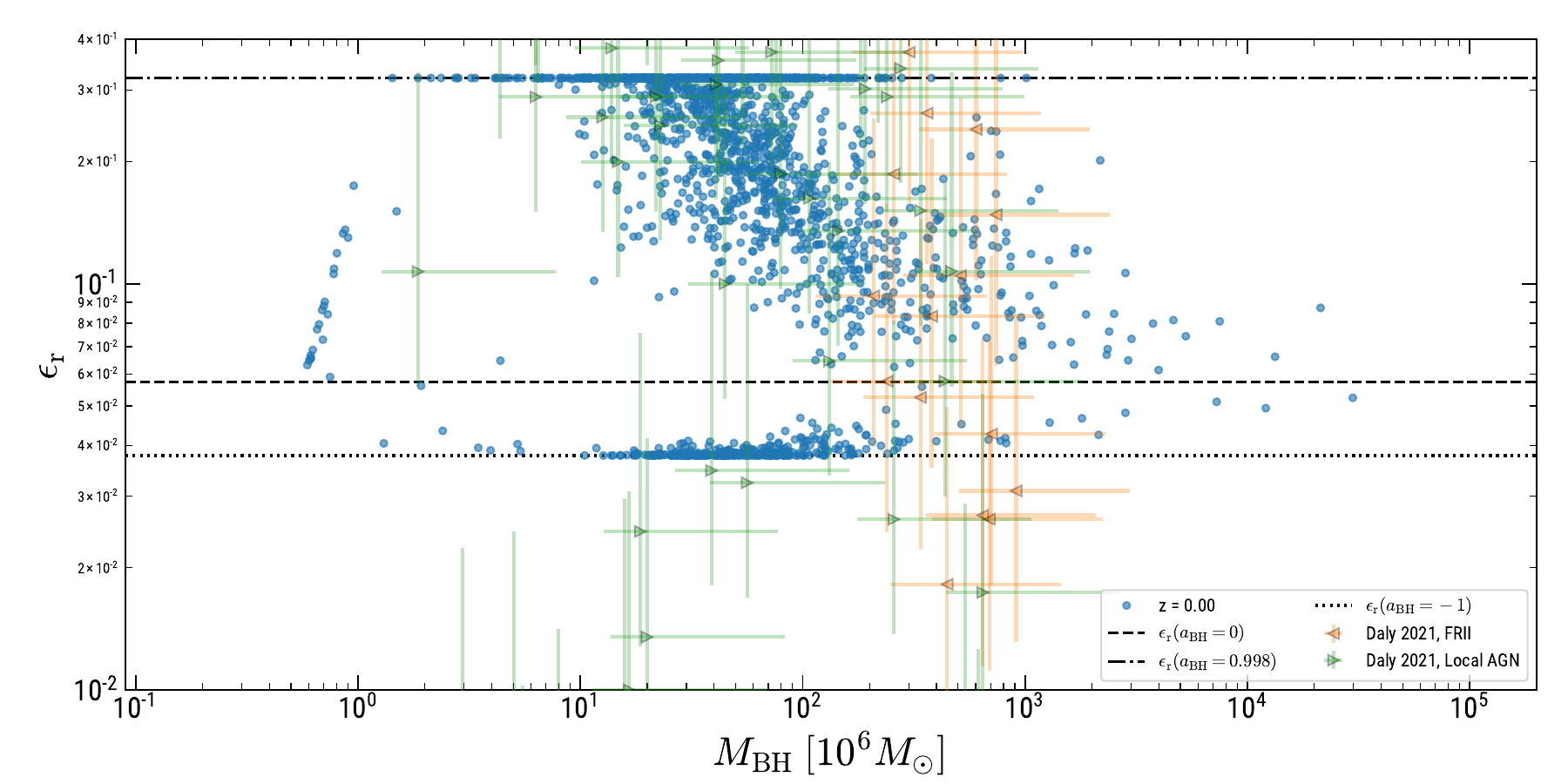}
    \caption{Radiative efficiencies of the BH populations at redshift $z=0$, as a function of mass, for the \textsc{Box4} run. The triangles show the collection of empirical estimates of the radiative efficiency by \citet{Daly2021}. The dotted, dashed and dash-dotted lines mark the values of the efficiency corresponding to $a=-1,0,0.998$, respectively, for reference. The population in the bottom part of the figure represents the BH that are accreting in counter-rotating conditions.}
    \label{fig:radeff_statistics}
\end{figure*}
Fig.~\ref{fig:radeff_statistics} shows the radiative efficiency $\minsub{\epsilon}{r}$ across the BH sample at $z=0$, as a function of $\minsub{M}{BH}$. The trends shown in Fig.~\ref{fig:spin_statistics_box} are reflected in the distribution of $\minsub{\epsilon}{r}$. Indeed, each BH has its own value of  $\minsub{\epsilon}{r}$ at each instant, dependent on $a$ (see Fig.~\ref{fig:spin_funcs}). We observe predominantly high efficiency ($\sim 0.32$) at intermediate BH masses ($10^{6}\lesssim\minsub{M}{BH}/\msun\lesssim 2 \times 10^{7}$). A lower efficiency value becomes more likely at higher masses (above $4\times 10^{7} \msun$), whereas at the highest masses (above $5\times 10^{8} \msun$) efficiencies have systematically lower values ($\sim 0.06-0.1$). The radiative efficiency also depends on whether, at a given instant, accretion is proceeding in co- or counter-rotating accretion conditions. Points with $\minsub{\epsilon}{r}$ below the dashed line correspond to BHs that are accreting in counter-rotating conditions. Indeed, the probability of having counter-rotating conditions  increases with mass above $4\times 10^{7} \msun$ (bottom panel of Fig.~\ref{fig:binned_disc_prop}). We also compare our simulated sample with a collection of radiative efficiency factors provided in \cite{Daly2021}. Since the simulation points refer to the sample at $z=0$, we consider all the sources in the observational catalogue that have $z<0.2$\footnote{Note that we exclude the sources catalogued as LINERs.}.

\section{Discussion}\label{sec:discussion}

\subsection{Evolution of the BH spin direction}
Our algorithm for spin evolution proceeds through accretion episodes that modify $\minsub{\bm{j}}{BH}$ as an effect of the external change in $\minsub{\bm{j}}{g}$. As shown in Fig.~\ref{fig:spin_tilt}, only if $\minsub{J}{d}\gg\minsub{J}{BH}$ a single accretion episode is able to induce complete alignment with $\minsub{\bm{j}}{d}$ and therefore with $\minsub{\bm{j}}{g}$. Conversely, if $\minsub{J}{d}\ll\minsub{J}{BH}$, then $\minsub{J}{BH}$ is insensitive to external change. We observe that in our simulations accretion episodes are in general characterised by $\minsub{J}{d}/\minsub{J}{BH}\lesssim 1$ (Fig.~\ref{fig:spin_tilt}), therefore accretion episodes induce only partial alignment of $\minsub{\bm{j}}{BH}$ with $\minsub{\bm{j}}{d}$. Larger values of $\minsub{J}{d}/\minsub{J}{BH}$ reduce the misalignment between $\minsub{\bm{j}}{BH}$ and $\minsub{\bm{j}}{g}$ to a larger degree. The relation between $\minsub{\bm{j}}{BH}$ and $\minsub{\bm{j}}{d}$ (and hence $\minsub{\bm{j}}{g}$) on timescales longer than a single accretion episode depends on $\minsub{J}{d}/\minsub{J}{BH}$ and on the variability of $\minsub{\bm{j}}{g}$. For the reference BH considered in \textsc{dfrogin}, the top panel of Fig.~\ref{fig:multiplot_asin} shows that before $t\sim2.6$ Gyr $\minsub{\bm{j}}{g}$ varies gradually with time, although with some variability on short (i.e. $\lesssim$ 1 Myr) timescales. On the other hand, $\minsub{\bm{j}}{g}$ changes erratically after $t\sim2.6$ Gyr. In the former case $\minsub{\bm{j}}{BH}$ manages to follow the average evolution of $\minsub{\bm{j}}{g}$, whereas the two directions are decoupled in the latter. The middle panel of Fig.~\ref{fig:binned_disc_prop} shows that statistically such large misalignment is more probable at the highest masses. Combined with low values of $\minsub{J}{d}/\minsub{J}{BH}$, it results in counter-rotating accretion conditions to be more frequent with increasing mass (bottom panel of Fig.~\ref{fig:binned_disc_prop}).

\subsection{Evolution of the BH spin magnitude}
The BH spin magnitude evolution via gas accretion is driven by the amount of accreted mass and by the radius of the ISCO (Eq.~\ref{eq:spinupdate}). Therefore, it evolves more rapidly in high accretion rate phases. Furthermore, counter-rotating accreting gas is characterised by a larger ISCO (bottom panel of Fig.~\ref{fig:spin_funcs}), thus a larger \am{} per unit mass. The same amount of accreted mass induces a larger (negative) change in $a$ than if it were accreted in co-rotating conditions. At fixed mass, if co- and counter-rotating episodes occurred in equal number, the net effect would actually be a decrease in $a$ \citep{Dotti+2013}. Whether there is a trend to increase or decrease the spin magnitude depends on the accretion rate and on how frequent co- or counter-rotating accretion is. Moreover, mergers also contributes to influence $a$.
In Figs.~\ref{fig:spin_statistics_zoomins} and \ref{fig:spin_statistics_box} we analyse the trends of $a$ with BH mass and observe that BHs with $\minsub{M}{BH}\lesssim2\times10^{7}\msun$ show systematic spin-up. This effect is attributed to the prevalence of co-rotating accretion conditions (bottom panel of Fig.~\ref{fig:binned_disc_prop}). Mergers do not occur in this range of masses (see Fig.~\ref{fig:magorrian_box}), hence the trend is exclusively due to gas accretion. For $\minsub{M}{BH}\gtrsim2\times10^{7}\msun$ we observe a wide distribution of $a$, indicating that the behaviour strongly depends on the detailed history, and there is no systematic behaviour. In this mass regime, several elements contribute to the trend: $i)$ the fraction of counter-rotating episodes increases with mass (bottom panel of Fig.~\ref{fig:binned_disc_prop}) but it remains generally lower than 50\%; $ii)$ co-rotating conditions and hence spin-up still occur; $iii)$ only the BHs in the highest mass bins ($\gtrsim5 \times 10^{9}\msun$) reach a fraction $\gtrsim0.5$ and exhibit systematically low spins ($a\simeq0.2-0.3$). We also note that above $\minsub{M}{BH}\gtrsim 10^{8}$ BHs undergo several mergers (Fig.~\ref{fig:magorrian_box}) and Eddington ratios are generally highly sub-Eddington (e.g. second to last panel of Fig.~\ref{fig:multiplot_asin}). Therefore, mergers significantly contribute to the variation of the spin. Moreover, mergers with misaligned directions tend to significantly decrease $a$ (e.g. fourth panel of Fig.~\ref{fig:multiplot_asin}). From Fig.~\ref{fig:spin_statistics_box}, we infer that mergers and a more likely counter-rotating accretion are responsible for the wider distribution of values of $a$.

In Fig.~\ref{fig:spin_statistics_box} we also note that BHs close to the seeding mass ($\minsub{M}{BH}\simeq5.5\times10^{5}\msun$) are characterised by a steep increase of $a$ with mass. The region in the $a-\minsub{M}{BH}$ plane that these BHs occupy is mostly determined by $\minsub{M}{BH,seed}$. However, the initial value of $a$ does not affect the following evolution, since any initial spin value is quickly evolved to maximal due to large accretion rates.

Overall, the trends in $a$ as a function of $\minsub{M}{BH}$ are consistent between the zoom-in simulations (Fig.~\ref{fig:spin_statistics_zoomins}) and the \textsc{Box4} (Fig.\ref{fig:spin_statistics_box}), which even have different resolutions. The distribution of $a$ is compatible with the observations within the uncertainties, across the entire mass range.

\subsection{Radiative efficiency}

The radiative efficiency plays an important role. First of all, it enters the computation of the Eddington accretion rate. Since the BH accretion rate cannot exceed this rate in our model, the efficiency directly affects the Eddington-limited growth phases. According to Eq.~\ref{eq:exp_growth}, the \textit{e}-folding timescale
\begin{equation}
    \minsub{\tau}{\minsub{M}{BH}}=\frac{\epsilon_{\rm r}\minsub{\tau}{S}}{\minsub{f}{Edd}(1-\epsilon_{\rm r})}
\end{equation}
depends on the efficiency. Eddington-limited co-rotating accretion on a maximally spinning BH (i.e. $\epsilon_{\rm r}\sim0.32$) implies $\minsub{\tau}{\minsub{M}{BH}}\sim 210$ Myr, whereas in counter-rotating conditions ($\epsilon_{\rm r}\sim0.038$) $\minsub{\tau}{\minsub{M}{BH}}\sim 18$ Myr. A lower efficiency leads to a significantly faster BH growth. In our simulations, Eddington-limited phases are characterised by maximally-spinning BHs and co-rotating accretion, thus growth proceeds with $\epsilon_{\rm r}\sim0.32$ (see e.g. $\epsilon_{\rm r}$ panel in Fig.~\ref{fig:multiplot_asin}).

The efficiency also controls the amount of feedback energy released to the surroundings, at a given $\somedot{M}{BH}$ (see Eq.~\ref{eq:feed_energy}). A lower efficiency implies less released energy and subsequent increased accretion. It also means a faster increase in $\minsub{M}{BH}$ (because of the factor $1-\minsub{\epsilon}{r}$ in Eq.~\ref{eq:mratio}). This in turn boosts $\somedot{M}{BH}$ (due to the $\propto\minsub{M}{BH}$ dependence in Eddington-limited phases or $\propto\minsub{M}{BH}^2$ otherwise, see Eq.~\ref{eq:bondi}), leading to stronger feedback outbursts that hinder accretion. In addition, since the Eddington accretion rate depends on the efficiency, the switch to maintenance mode feedback is also affected. Overall, the effects just discussed contribute in a non-trivial way to modify the evolutionary path of a BH through the feedback loop. 
Our simulations -- where all these processes are self-consistently taken into account -- show that BHs are on the observed correlation between BH mass and stellar mass (Figs.~\ref{fig:magorrian_box} and \ref{fig:magorrian_zoomins}). While the detailed evolutionary path along the plane $\minsub{M}{BH}-\minsub{M}{*}$ changes, each BH is eventually able to approach the correlation, implying that the BHs still grow in an overall self-regulated scenario.

Fig.~\ref{fig:radeff_statistics} highlights that $\minsub{\epsilon}{r}$ tends to decrease with increasing mass at the high-mass end. We also find that the distribution of our simulated sample is compatible with the distribution of empirical estimates by \cite{Daly2021}, within the uncertainties. Note that the empirical sample is obtained with a method that does not rely on a specific accretion disc model. Therefore, while our sample has an upper and lower limit for the efficiency determined by our theoretical assumption on the disc structure, the interpretation of the empirical estimates is not bound to such an assumption. 

\section{Comparison with previous works}\label{sec:previous_work_comparison}

\cite{Dubois+2014} and \cite{Bustamante&Springel2019} perform a statistical study similar to ours. The former simulate a cosmological volume with size and resolution comparable to \textsc{Box4} and the latter a smaller box with higher resolution ($\minsub{L}{box}=25\;h^{-1}\;\rm cMpc$ and $\minsub{m}{DM}=8.4\times10^{6}\msun$). Our spin evolution model follows \cite{Dubois+2014}, although we track the spin on-the-fly whereas in their work the spin is tracked in post-processing. $\minsub{\epsilon}{r}$ is assumed to be fixed to 0.1. They include a thermal feedback channel for $\minsub{f}{Edd} > 0.01$ and a bipolar outflow launched in the direction of the local gas otherwise, but both do not depend on $a$.
\cite{Bustamante&Springel2019} adopt an on-the-fly spin update algorithm similar to ours, with a variable $\minsub{\epsilon}{r}$ affecting the thermal feedback mode. However, below a mass-dependent $\minsub{f}{Edd}$ threshold, BHs inject purely kinetic energy, with a fixed efficiency and random direction, isotropic on average. Moreover, when the sub-grid accretion disc is affected by self-gravity the \am{} direction of each accretion episode is extracted from a chosen angular distribution. In our implementation we assume it is fixed and equal to the gas \am{} direction at all times.
Interestingly, the trends we find in Fig.~\ref{fig:spin_statistics_box} are in agreement with both works, regardless of whether or not $\minsub{\epsilon}{r}$ depends on $a$ or whether the spin is evolved in post-processing rather than on-the-fly. This indicates that the variability of the radiative efficiency does not play a significant role in setting these trends. On the other hand, $\minsub{\epsilon}{r}$ does affect AGN feedback and BH growth (Sec.~\ref{sec:agn_feedback}). As a result, a different prescription for $\minsub{\epsilon}{r}$ changes the detailed time evolution of BH mass and stellar mass, leading to a different path towards the $\minsub{M}{BH}-\minsub{M}{*}$ relation and more scatter around it \citep{Bustamante&Springel2019}. Nonetheless, Fig.~\ref{fig:magorrian_box} shows that the self-regulated scenario is still present and eventually leads the BHs on the relation.

We find that the dynamical state of the feeding material, combined with the key parameter $\minsub{J}{d}/\minsub{J}{BH}$, plays a significant role in the evolution of the BH spin due to gas accretion (Fig.~\ref{fig:binned_disc_prop} and \ref{fig:spin_tilt}). 
\cite{Dotti+2013} conclude that when $\minsub{J}{d}/\minsub{J}{BH}\ll 1$ (as it occurs in our simulations, see Fig.~\ref{fig:binned_disc_prop}), the feeding \am{} distribution determines whether $a$ increases or decreases. 
Our findings regarding the behaviour of the spin in response to the feeding conditions are in agreement with theirs, as well as with \cite{Dubois+2014} and \cite{Bustamante&Springel2019}. When the gas \am{} direction varies slowly and its misalignment with respect to the BH spin is small (indicating a preferential direction), the BHs are maximally spinning. This generally occurs at $\minsub{M}{BH}\lesssim 10^{8}\msun$ (see  Fig.~\ref{fig:multiplot_asin}, \ref{fig:10mostmassive_box}, \ref{fig:binned_disc_prop}).  Conversely, uncorrelated gas \am{} directions lead to increasingly probable counter-rotating accretion and BH spin-down (bottom panel of Fig.~\ref{fig:binned_disc_prop}, Sec.~\ref{sec:cosmo_box_results}). 
We note that the large scatter in $a$ at high masses (Fig.~\ref{fig:spin_statistics_box}) is found also by \cite{Dubois+2014a} and \cite{Bustamante&Springel2019}, despite the different prescriptions for AGN feedback. This might indicate that its effect is to generally induce loss of coherence in the \am{} distribution (middle panel of Fig.~\ref{fig:binned_disc_prop}), regardless of the specific channel.

In contrast to our model, \cite{Bustamante&Springel2019} introduce stochasticity in $\minsub{\bm{j}}{d}$ (Eq.~\ref{eq:disc_dir}) at the sub-resolution level, on top of the resolved variability in the simulation. They assume that in the self-gravity regime $\minsub{\bm{j}}{d}$ is extracted from a distribution that ranges from random (isotropic) to concentrated around the preferential axis set by the local gas \am{} (anisotropic). As a result, they observe a more pronounced decoupling between $\minsub{\bm{j}}{d}$ and $\minsub{\bm{j}}{BH}$ compared to us, in case of an isotropic distribution. On the other hand, they find the same widening of the distribution of $a$ at high BH masses, regardless of the degree of anisotropy.
\cite{Dubois+2014a} also explore the effect of varying the distribution of $\minsub{\bm{j}}{d}$ at the sub-resolution level. However, they introduce stochasticity in all accretion regimes. They find that even a slight anisotropy leads to results that are very similar to the completely coherent case (i.e. when the gas preserves the \am{} direction as measured by the simulation). Only if the gas \am{} is random oriented at all masses, then BHs settle on $a\sim 0.2-0.3$ \citep{King+2008} and an increasing trend of spin with mass is found \citep{Fanidakis+2011} (in those conditions the trend is produced by mergers, that tend to bring slowly-spinning BHs to values around $0.7$ for $\minsub{M}{BH}\gtrsim 10^{9} \msun$).
One possibility to explain these results is that when gas accretion is sub-dominant, the trend is mainly driven by mergers. 
However, we observe that above $\minsub{M}{BH}\gtrsim 10^{8}\msun$ gas accretion does contribute to spin evolution, inducing both spin-up and spin-down (Fig.~\ref{fig:10mostmassive_box}). The net effect depends crucially on the accretion rate, the BH environment and the level of anisotropy of the feeding gas. This level varies widely over time and across the BH population. Mergers also contribute to modify the spin value, but the relative contribution of gas accretion and mergers to spin evolution is different depending on the detailed cosmological history. We postpone a systematic study to a future work. 

Finally, we note that we do not integrate the full differential equation that describes the precession and alignment process, due to the coupling between the gas distribution in the accretion disc and the BH spin \citep[at variance with, e.g.,][]{Fiacconi+2018}. Rather, we assume that the process is discretised in accretion episodes and is globally taken into account, while the total \am{} is conserved (Sec.~\ref{sec:spinevol_algorithm}), following \cite{King+2005}. \cite{Fiacconi+2018} developed an algorithm that includes the full detailed treatment, although it requires high temporal and spatial resolution. In fact, timesteps as low as $10^{-3}$ Myr are required in some cases to meaningfully integrate the differential equation, which are prohibitive in a full cosmological context. Moreover, their model measures directly the inflow properties as resolved by the simulation at the sub-resolution boundary, rather than using an effective prescription such as the Bondi parametrisation. Therefore, it is suitable for high-resolution simulations (e.g. $\sim$ pc scale). 
In contrast, we assume that the mass rate through the sub-resolution accretion disc is equal to the mass accretion rate onto the BH and it is identical to the Bondi accretion rate at all times. We further note that \cite{Fiacconi+2018} model spin evolution only due to gas accretion, although mergers are expected to contribute to spin evolution for $\minsub{M}{BH}\gtrsim 10^{8}\msun$ \citep{Fanidakis+2011,Dubois+2014a}. Although they do not produce a statistical sample of BHs, they perform a suite of simulations aimed at mimicking a range of realistic conditions. Despite the approach is different in a number of numerical aspects, the expected effect of the gas accretion channel on spin evolution is in line with ours: systematic spin-up for $\minsub{M}{BH}\lesssim 10^{7}\msun$ and a wider distribution of $a$ at higher masses. 

\section{Conclusions}\label{sec:conclusion}

We implement a sub-resolution model to track the evolution of BH spins due to gas accretion and mergers in large scale cosmological simulations. The model assumes the presence of a misaligned thin accretion disc perturbed by the metric of a spinning BH, which in turn experiences a torque that modifies its spin direction. The BH radiative efficiency and Eddington accretion rate are dependent on the BH spin and thus variable across cosmic time. Their impact on accretion and feedback is therefore captured self-consistently. We design a simulation suite featuring idealised, isolated systems (Sec.~\ref{sec:ideal_gal_results} and \ref{sec:ideal_merger_results}) to validate the model and cosmological setups (Sec.~\ref{sec:zoomin_results} and \ref{sec:cosmo_box_results}) to investigate statistical properties of the BH population. We summarise our findings as follows.

\begin{itemize}
    \item The ability of a single accretion episode to modify the BH spin depends on the amount of mass and \am{} it carries with respect to the BH ($\minsub{J}{d}/\minsub{J}{BH}$, Fig.~\ref{fig:spin_tilt}). An accretion episode with larger $\minsub{J}{d}/\minsub{J}{BH}$ induces the BH spin direction $\minsub{\bm{j}}{BH}$ to tilt more towards the gas \am{} direction $\minsub{\bm{j}}{g}$.
    \item The evolution of the direction and the magnitude of the spin are tightly coupled. When $\minsub{J}{d}/\minsub{J}{BH}\lesssim 1$ (Fig.~\ref{fig:binned_disc_prop}), the feeding distribution of the gas \am{} directions determines whether $a$ preferably increases or decreases. If accretion occurs consistently along the same plane (e.g. Fig.~\ref{fig:isol_gal_accr}), spin-up is expected. Conversely, if $\minsub{\bm{j}}{g}$ changes direction erratically, counter-rotating conditions and spin-down can occur (Fig.~\ref{fig:multiplot_asin} and \ref{fig:binned_disc_prop}). 
    \item In a cosmological context, we identify two regimes, depending on the distribution of $a$ with BH mass (Fig.~\ref{fig:spin_statistics_box}). BHs with $\minsub{M}{BH}\lesssim 2 \times 10^7 \msun$ tend to be highly spinning ($a\gtrsim0.85$). At the high-mass range ($\minsub{M}{BH}\gtrsim 2 \times 10^{7}\msun$), $a$ exhibits a broad range of values.
    \item We observe a wide variety of evolutionary histories of $a$ (Fig.~\ref{fig:10mostmassive_box}), depending on the dynamical state of the gas feeding the BH and the occurrence of coalescences. This indicates that the level of anisotropy of $\minsub{\bm{j}}{g}$ and the relative contribution of mergers and accretion varies across the BH population.
    \item When $\minsub{\bm{j}}{g}$ exhibits some degree of coherence and varies slowly (generally at $z\gtrsim2$ and $\minsub{M}{BH}\lesssim 2 \times 10^7 \msun$), $\minsub{\bm{j}}{BH}$ follows the average evolution of $\minsub{\bm{j}}{g}$ and small misalignment is observed (Fig.~\ref{fig:binned_disc_prop}). At late times ($z\lesssim2$), $\minsub{\bm{j}}{g}$ shows large and abrupt changes, while $\minsub{\bm{j}}{BH}$ is stable over long periods (hundreds of Myr, e.g. Fig.~\ref{fig:multiplot_asin}). Indeed, since $\minsub{J}{d}/\minsub{J}{BH}\lesssim 1$, accretion episodes are not able to modify significantly $\minsub{\bm{j}}{BH}$. This frequently leads to large misalignment.
    \item The tendency for maximal spin in the low-mass range is due to accretion of co-rotating gas with small misalignment (Fig.~\ref{fig:binned_disc_prop}). The wide range of values of $a$ in the high-mass range is due to mergers and more isotropically distributed $\minsub{\bm{j}}{g}$, this results in a probability of counter-rotating accretion that increases with mass (Figs.~\ref{fig:10mostmassive_box} and \ref{fig:binned_disc_prop}). The distribution of $\minsub{\bm{j}}{g}$ arises self-consistently, as measured from the simulation. 
    \item Including a self-consistent $\minsub{\epsilon}{r}$ has an important effect in determining the BH growth rate in the Eddington-limited phases. A higher efficiency during these phases implies a lower growth rate. Our statistical sample shows that BHs with $\minsub{M}{BH}\lesssim 4 \times 10^7 \msun$ have efficiencies always around $0.32$, whereas the most massive BHs have generally lower efficiencies (Fig.~\ref{fig:radeff_statistics}).
    \item Although the variable $\minsub{\epsilon}{r}$ modifies the detailed path on the $\minsub{M}{BH}-\minsub{M}{*}$ plane, BHs eventually approach the observed correlation, indicating self-regulated growth.
    \item The spin with which BHs are initialised is erased quickly after they are seeded. Massive BHs can retain some information on the dynamical state of the gas they recently accreted (Figs.~\ref{fig:10mostmassive_box} and \ref{fig:binned_disc_prop}).
\end{itemize}

We caution that so far we have coupled the spin evolution model to a single channel of energy injection, namely purely thermal, aimed at reproducing the radiative feedback. An additional mechanism, in which large-scale jets (10s-100s of kpc) are the key actors, is thought to be specifically relevant at the high-mass end. We plan to include this feature in future works, by coupling the spin evolution model to a feedback channel from jets, dependent on the spin magnitude and direction. 

\begin{acknowledgements}
    LS acknowledges support from `BiD4BEST' - European Innovative Training Network (ITN) funded by the Marie Sk\l{}odowska-Curie Actions (860744) in Horizon 2020. 
    MV is supported by the Fondazione ICSC, Spoke 3 Astrophysics and Cosmos Observations, National Recovery and Resilience Plan (PNRR) Project ID CN\_00000013 "Italian Research Center on High-Performance Computing, Big Data and Quantum Computing" funded by MUR (Missione 4, Componente 2, Investimento 1.4: Potenziamento strutture di ricerca e creazione di campioni nazionali di R\&S, M4C2-19) - Next Generation EU. MV also acknowledges partial financial support from the INFN Indark Grant.
    This work was supported by the Deutsche Forschungsgemeinschaft (DFG, German Research Foundation) under Germany’s Excellence Strategy - EXC-2094 - 390783311 and we are especially grateful for the support through the Computational Center for Particle and Astrophysics (C2PAP). Some of the analyses reported in this work have been performed using \textsc{pynbody} \citep{Pontzen+2013}
\end{acknowledgements}

%
\bibliographystyle{aa} 
\bibliography{biblio} 
%

\end{document}